\documentclass[twocolumn,trackchanges]{aastex701}

\usepackage{xspace}
\usepackage[T1]{fontenc}

\newcommand{\chandra}{{\it Chandra}\xspace}

\newcommand{\nicer}{{\it NICER}\xspace}
\newcommand{\suz}{{\it Suzaku}\xspace}
\newcommand{\xmm}{{\it XMM-Newton}\xspace}

\newcommand{\nustar}{\textit{NuSTAR}\xspace}

\newcommand{\hxmt}{{\it HXMT}\xspace}

\newcommand{\microsec}{\ensuremath{\rm \mu s}\xspace}
\newcommand{\asec}{\ensuremath{^{\prime\prime}}\xspace}
\newcommand{\amin}{\ensuremath{^{\prime}}\xspace}

\newcommand{\msun}{\ensuremath{M_{\odot}}\xspace}

\newcommand{\figref}{Figure~\ref}

\shorttitle{IACHEC Timing cross-cal with the Crab pulsar}
\shortauthors{Bachetti et al.}

\begin{document}

\title{A simple, flexible method for timing cross-calibration of space missions}

\author[0000-0002-4576-9337]{Matteo Bachetti}
\affiliation{INAF-Osservatorio Astronomico di Cagliari,
via della Scienza 5
I-09047 Selargius (CA), Italy}
\email{matteo.bachetti@inaf.it}
\correspondingauthor{Matteo Bachetti, matteo.bachetti@inaf.it}

\author[0000-0002-2359-1857]{Yukikatsu Terada}
\affiliation{Graduate School of Science and Engineering, Saitama University, 255 Shimo-Okubo, Sakura-ku, Saitama-shi Saitama 338-8570, Japan}
\affiliation{Institute of Space and Astronautical Science( ISAS) , Japan Aerospace Exploration Agency( JAXA) , 3-1-1 Yoshinodai, Chuo-ku, Sagamihara, Kanagawa 252-5210, Japan}
\email{terada@mail.saitama-u.ac.jp}

\author[0000-0001-8195-6546]{Megumi Shidatsu}
\affiliation{Ehime University, Graduate School of Science and Engineering, 2-5, Bunkyo-cho, Matsuyama, Ehime, 790-8577, Japan}
\email{shidatsu.megumi.wr@ehime-u.ac.jp}

\author[0000-0001-9803-3879]{Craig B. Markwardt}
\affiliation{X-ray Astrophysics Laboratory, NASA Goddard Space Flight Center Code 662, Greenbelt, MD, 20771, USA}
\email{craigm@milkyway.gsfc.nasa.gov}

\author{Yong Chen}
\affiliation{Key Laboratory of Particle Astrophysics,
Institute of High Energy Physics, Chinese Academy of Science, 19B Yuquan Road, Shijingshan District, Beijing, 100049, China}
\email{ychen@ihep.ac.cn}

\author{Weiwei Cui}
\affiliation{Key Laboratory of Particle Astrophysics,
Institute of High Energy Physics, Chinese Academy of Science, 19B Yuquan Road, Shijingshan District, Beijing, 100049, China}
\email{cuiww@ihep.ac.cn}

\author[0000-0002-8151-1990]{Giancarlo Cusumano}
\affiliation{INAF - Istituto di Astrofisica Spaziale e Fisica Cosmica di Palermo, Via U. La Malfa 153, 90146 Palermo, Italy}
\email{giancarlo.cusumano@inaf.it}

\author{Dawei Han}
\affiliation{Key Laboratory of Particle Astrophysics,
Institute of High Energy Physics, Chinese Academy of Science, 19B Yuquan Road, Shijingshan District, Beijing, 100049, China}
\email{dwhan@ihep.ac.cn}

\author[0000-0002-5203-8321]{Shumei Jia}
\affiliation{Key Laboratory of Particle Astrophysics,
Institute of High Energy Physics, Chinese Academy of Science, 19B Yuquan Road, Shijingshan District, Beijing, 100049, China}
\email{jiasm@ihep.ac.cn}

\author{Chulsoo Kang}
\affiliation{Graduate School of Science and Engineering, Saitama University, 255 Shimo-Okubo, Sakura-ku, Saitama-shi Saitama 338-8570, Japan}
\email{m809001y@mails.cc.ehime-u.ac.jp}

\author[0000-0002-3869-7996]{Vinay L.\ Kashyap}
\affiliation{Center for Astrophysics | Harvard \& Smithsonian, Cambridge, MA 02138, USA}
\email{vkashyap@cfa.harvard.edu}

\author[0000-0002-7889-6586]{Lucien Kuiper}
\affiliation{SRON - Space Research Organisation Netherlands, Niels Bohrweg 4, 2333 CA, Leiden, The Netherlands}
\email{l.m.kuiper@sron.nl}

\author[0000-0003-4585-589X]{Xiaobo Li}
\affiliation{Key Laboratory of Particle Astrophysics,
Institute of High Energy Physics, Chinese Academy of Science, 19B Yuquan Road, Shijingshan District, Beijing, 100049, China}
\email{lixb@ihep.ac.cn}

\author{Yugo Motogami} 
\affiliation{Graduate School of Science and Engineering, Saitama University, 255 Shimo-Okubo, Sakura-ku, Saitama-shi Saitama 338-8570, Japan}
\email{y.motogami.738@ms.saitama-u.ac.jp}

\author{Naoyuki Ota}
\affiliation{RIKEN Nishina Center, 2-1 Hirosawa, Wako, Saitama 351-0198, Japan}
\affiliation{Tokyo University of Science, 1-3 Kagurazaka, Shinjuku, Tokyo 162-8601, Japan}
\email{naoyuki.ota@riken.jp}

\author[0009-0004-5622-1854]{Simone Pagliarella}
\affiliation{INAF Istituto di Astrofisica e Planetologia Spaziali, Via del Fosso del Cavaliere 100, 00133 Roma, Italy}
\affiliation{Tor Vergata University of Rome, Via Della Ricerca Scientifica 1, 00133 Roma, Italy}
\affiliation{Dipartimento di Fisica, Università degli Studi di Roma ``La Sapienza'', P.le Aldo Moro 2, 00133 Roma, Italy}
\email{simone.pagliarella@inaf.it}

\author[0000-0002-4656-6881]{Katja Pottschmidt}
\affiliation{X-ray Astrophysics Laboratory, NASA Goddard Space Flight Center Code 662, Greenbelt, MD, 20771, USA}
\affiliation{University of Maryland, Baltimore County, 1000 Hilltop Circle, Baltimore, Maryland, 21250, United States}
\altaffiliation{Deceased 17 June 2025}
\email{katja.pottschmidt-1@nasa.gov}

\author{Simon R. Rosen}
\affiliation{Serco for the European Space Agency (ESA), European Space Astronomy Centre, Camino Bajo del Castillo s/n, E-28692 Villanueva de la Cañada, Madrid, Spain}
\email{simon.rosen@ext.esa.int}

\author[0000-0003-2377-2356]{Arnold Rots}
\affiliation{Center for Astrophysics | Harvard \& Smithsonian, Cambridge, MA 02138, USA}
\email{arots@cfa.harvard.edu}

\author[0000-0003-2008-6887]{Makoto Sawada} 
\affiliation{Department of Physics, Rikkyo University, Tokyo 171-8501, Japan}
\email{makoto.sawada@rikkyo.ac.jp}

\author[0000-0002-1190-0720]{Mutsumi Sugizaki}
\affiliation{Advanced Research Center for Space Science and Technology, Kanazawa University, Kakuma, Kanazawa, Ishikawa, 920-1192, Japan}
\email{sugizaki@se.kanazawa-u.ac.jp}

\author{Toshihiro Takagi}
\affiliation{Ehime University, Graduate School of Science and Engineering, 2-5, Bunkyo-cho, Matsuyama, Ehime, 790-8577, Japan}
\email{takagi.toshihiro.bb@ehime-u.ac.jp}

\author{Takuya Takahashi}
\affiliation{RIKEN Nishina Center, 2-1 Hirosawa, Wako, Saitama 351-0198, Japan}
\affiliation{Tokyo University of Science, 1-3 Kagurazaka, Shinjuku, Tokyo 162-8601, Japan}
\email{1225545@ed.tus.ac.jp}

\author[0000-0002-8801-6263]{Toru Tamagawa}
\affiliation{RIKEN Pioneering Research Institute, 2-1 Hirosawa, Wako, Saitama 351-0198, Japan}
\affiliation{RIKEN Nishina Center, 2-1 Hirosawa, Wako, Saitama 351-0198, Japan}
\affiliation{Tokyo University of Science, 1-3 Kagurazaka, Shinjuku, Tokyo 162-8601, Japan}
\email{tamagawa@riken.jp}

\author[0000-0003-3127-0110]{Youli Tuo}
\affiliation{Institut f\"{u}r Astronomie und Astrophysik, Kepler Center for Astro and Particle Physics, Eberhard Karls Universit\"{a}t T\"{u}bingen, Sand 1, 72076 T\"{u}bingen, Germany}
\email{youli.tuo@astro.uni-tuebingen.de}

\author[0000-0001-9108-573X]{Yi-Jung Yang}
\affiliation{Center for Astrophysics and Space Science (CASS), New York University Abu Dhabi, PO Box 129188, Abu Dhabi, UAE}
\email{yjyang312@gmail.com}

\author[0009-0005-0819-0819]{Marina Yoshimoto}
\affiliation{Ehime University, Graduate School of Science and Engineering, 2-5, Bunkyo-cho, Matsuyama, Ehime, 790-8577, Japan}
\email{yoshimoto.marina.gr@ehime-u.ac.jp}

\author[0000-0001-8869-0672]{Juan Zhang}
\affiliation{Key Laboratory of Particle Astrophysics,
Institute of High Energy Physics, Chinese Academy of Science, 19B Yuquan Road, Shijingshan District, Beijing, 100049, China}
\email{zhangjuan@ihep.ac.cn}

\begin{abstract}
The timing (cross-)calibration of astronomical instruments is often done by comparing pulsar times-of-arrival (TOAs) to a reference timing model. In high-energy astronomy, the choice of solar system ephemerides and source positions used to barycenter the photon arrival times has a significant impact on the procedure, requiring a full reprocessing of the data each time a new convention is used.
Our method, developed as part of the activities of the International Astronomical Consortium for High Energy Calibration (IACHEC), adapts an existing pulsar solution to arbitrary JPL ephemerides and source positions by simulating geocentric TOAs and refitting timing models (implemented with PINT). We validate the procedure and apply it to thousands of observations of the Crab pulsar from 15 missions spanning 1996--2025, demonstrating inter-ephemeris TOA consistency at the $\lesssim5 \microsec$ level, using the DE200/FK5-based Jodrell Bank Monthly Ephemeris as a common reference. We release the TOAExtractor open-source tool and a TOA database to support future calibration and scientific studies. Instrument timing performance is broadly consistent with mission specifications; the X-ray-to-radio phase offset varies with energy and time at a level that is marginally consistent with the uncertainties of the radio ephemeris, motivating coordinated
multiwavelength follow-up.
\end{abstract}


\section{Introduction} \label{sec:intro}

The International Astronomical Consortium for High Energy Calibration (IACHEC) is a collaboration of space missions that aims to improve the calibration of high-energy instruments.
IACHEC brings together scientists and calibration teams from various observatories to share expertise, cross-validate results, and develop best practices for calibration in X-ray and gamma-ray astronomy.
Through coordinated efforts, IACHEC facilitates the comparison of data from different missions, identifies systematic discrepancies, and works towards establishing common calibration standards.
This collaborative approach is essential for ensuring the reliability and consistency of high-energy astrophysical measurements across the international community.

The Timing Working Group (TWG) focuses on the temporal cross-calibration of different missions.
The goal of the TWG is to develop methods to compare the timing performance of different missions, using pulsars as a common clock.
Pulsars are neutron stars (NSs) that emit beams of radiation swiping the sky like a lighthouse (giving the characteristic pulsed signal).
Discovered in the radio band \citep{hewishObservationRapidlyPulsating1968}, pulsars have later been detected at all wavelengths, from radio to gamma-rays \citep{lovelacePulsarNP05321968,tananbaumDiscoveryPeriodicPulsating1972,nelsonOpticalTimingPulsar1970,guptaHighenergyPulsedGamma1978,abdoSecondFermiLarge2013}.

With the mass of 1-2\,\msun and a radius of about 10\,km, NSs are the densest objects in the universe excluding stellar-mass black holes.
Being so compact, they are effectively point-like as seen from nearby stars, and tidal forces are usually negligible.
Their rotational speed can change over time due to a number of intrinsic and external factors. The emission of electromagnetic radiation and a particle wind from their rotating magnetic field produces a characteristic spin-down.

NSs, however, are probably not homogeneous spheres, but they likely have a complex internal structure, with a solid crust and a superfluid core.
``Glitches'' in their spin evolution, that produce sudden changes of spin period or its derivatives, are thought to arise exactly from the pinning and unpinning of the internal layers of the star \citep[see][for a review]{manchesterPulsarGlitches2018}.
Accretion from a companion star can also change the spin period of a pulsar, either by adding angular momentum to the star (spin-up) or by removing it (spin-down) \citep{ghoshDiskAccretionMagnetic1978}.
The fastest-rotating pulsars, known as millisecond radio pulsars (MSPs), are usually old pulsars that have gone through a spin-up process through accretion and have a low magnetic field, which limits the rotational noise and further stabilizes the rotation. In fact, some MSPs have rotations that can be predicted with a precision of microseconds in a few years \citep[e.g.][]{reardonNeutronStarMass2024}.

Even though the precise emission mechanisms of pulsars are still debated, NSs have a very strong surface magnetic field (from 10$^{8}$ to $\sim$10$^{15}$\,G), which is responsible for most of their phenomenology. First of all, the energy loss from the rotating dipole field that slows pulsars down is also believed to power most radio pulsars (for this reason, also called rotation-powered pulsars, or RPPs) with $B\lesssim10^{12}$\,G \citep{lorimerBinaryMillisecondPulsars2008}.
Accreting pulsars are instead powered by the accretion of matter which is funneled on the surface by the magnetic field \citep{ghoshDiskAccretionMagnetic1978}, producing a bright X-ray emission.
The highest-magnetic-field pulsars, known as magnetars, are believed to be powered by the decay of the magnetic field itself, and they show a complicated phenomenology, including giant flares and bursts at high energies \citep{reaMagnetars2025}.

For their high rotational stability, pulsars have often been used as standard clocks, with applications ranging from the detection of gravitational waves \citep{taylorNewTestGeneral1982,agazieNANOGrav15Yr2023,eptacollaborationSecondDataRelease2023} to space navigation \citep{sheikhSpacecraftNavigationUsing2006,emadzadehXRayPulsarBasedRelative2011,andersonValidationPulsarPhase2015,mitchellSextantXRayPulsar2018, yidiReviewXrayPulsar2023}.
Even more intriguing, pulsars have been proposed as reference clocks to establish a new time scale, independent of atomic clocks \citep[e.g.][]{hobbsPulsarbasedTimescaleInternational2020}.
Another application, which is the subject of this work, has been the timing calibration of astronomical instruments \citep[e.g.][]{kuiperAbsoluteTimingIBIS2003,rotsAbsoluteTimingCrab2004,smithPulsarTimingFermi2008,teradaInOrbitTimingCalibration2008,molkovAbsoluteTimingCrab2010,martin-carrilloRelativeAbsoluteTiming2012,cusumanoTimingAccuracySwift2012,denevaLargeHighprecisionXRay2019,bachettiTimingCalibrationNuSTAR2021,basuAbsoluteTimeCalibration2021,tuoInorbitTimingCalibration2022,cusumanoTrackingLongtermTiming2024}.

PSR B0531+21, also known as PSR J0534+2200 or the Crab pulsar due to its position at the center of the Crab nebula M1, is a RPP with a spin period of $P\sim 33$\,ms, and a spin-down rate of $\dot{P} \sim 4.2 \times 10^{-13}$\,s\,s$^{-1}$, which corresponds to a spin-down luminosity of $\dot{E}=4\pi^2I\dot{P}/P^3\sim 4 \times 10^{38}$\,erg\,s$^{-1}$ \citep{lorimerBinaryMillisecondPulsars2008}.
Despite being young%
\footnote{The Crab nebula has been identified as the remnant of SN 1054, recorded in 1054 A.D. by astronomers all around the globe \citep{mayallCrabNebulaProbable1939}}
and not particularly stable by pulsar standards (including glitches and a variable spin down rate that amount to a large ``timing noise''), it is one of the best studied RPPs in the sky, due to being very bright across the electromagnetic spectrum and, not less importantly, due to the Crab nebula's relatively stable (and power law-like in the X-rays) spectrum which made it an excellent flux calibrator.
This provides pulsar astronomers with a wealth of ``free'' observations of the pulsar.
The pulsar's spin evolution has been tracked for four decades by the Jodrell Bank Observatory \citep{lyneTwentyThreeYearsCrab1993}, which still today provides a monthly ephemeris (the Jodrell Bank Monthly Ephemeris, hereafter JBE) of the pulsar's spin frequency.

This makes the Crab pulsar an excellent tool for tracking the timing calibration of X-ray instruments, and it has been used as such by many missions \citep[e.g.][]{kuiperAbsoluteTimingIBIS2003,rotsAbsoluteTimingCrab2004,smithPulsarTimingFermi2008,teradaInOrbitTimingCalibration2008,molkovAbsoluteTimingCrab2010,cusumanoTimingAccuracySwift2012,martin-carrilloRelativeAbsoluteTiming2012,bachettiTimingCalibrationNuSTAR2021,basuAbsoluteTimeCalibration2021,tuoInorbitTimingCalibration2022,cusumanoTrackingLongtermTiming2024}.

Typically, these missions use the JBE to fold the pulsar data and obtain pulse profiles and TOAs, which are then compared to the expected TOAs from the ephemerides.
The main problem with this approach is that the Jodrell Bank ephemerides are based on the DE200 solar system ephemeris \citep{standishOrientationJPLEphemerides1982}, which is quite outdated, and they do not account for proper motion, so assuming that the pulsar's position is fixed in the sky.
This makes sense to provide a consistent timing solution for the Crab pulsar, but it is not ideal for cross-calibration purposes, as it forces all missions to use the same ephemeris and source position, losing flexibility when gathering data from different missions, and often needing to perform separate data processing for timing calibration and scientific analysis.

In this paper, we present a new method for comparing the times of arrival (TOAs) of pulsar signals obtained with different instruments, even when they use different JPL ephemerides, source positions, and reference frames.
We apply the method to the temporal cross-calibration of different missions, using the Jodrell Bank Monthly Ephemeris of the Crab pulsar as a common reference.
In section~\ref{sec:process}, we describe the typical pulsar timing workflow and the specific data processing steps for X-ray and gamma-ray data.
In section~\ref{sec:jpleph}, we describe our method for adapting pulsar timing solutions to different ephemerides and source positions.
In section~\ref{sec:results}, we present the results of applying our method to a large dataset of Crab pulsar observations from multiple missions.
Finally, in section~\ref{sec:conclusions}, we summarize our findings and discuss the implications for future timing calibration efforts.
Appendix~\ref{sec:timedef} contains definitions of time scales and conventions used in pulsar timing, Appendix~\ref{sec:missiondataproc} contains details on the data processing for each mission used in this work.

\section{Processing} \label{sec:process}

\subsection{Typical pulsar timing workflow}
The pulsar signals are periodic in nature, repeating at the pulsar's spin period, which is typically in the range of milliseconds to seconds (with a few cases of pulsars with spin periods up to hours, see the reviews \citealt{lorimerBinaryMillisecondPulsars2008,reaMagnetars2025}).
It is common to express the signal at a time $t$ in terms of the pulsar's spin phase\footnote{We express the phase from 0 to 1. Other works might use 0-$2\pi$.} $\phi(t)$, defined through the pulsar's spin frequency $\nu$ and its time derivative $\dot{\nu}$, $\ddot{\nu}$ with the following Taylor expansion:
\begin{equation}\label{eq:phase}
\phi(t) = \nu(t-t_0) + \frac{1}{2}\dot{\nu}(t-t_0)^2 + \frac{1}{6}\ddot{\nu}(t-t_0)^3 + \ldots
\end{equation}
where $t_0$ is a reference TOA.

The single pulses of pulsars are typically too faint to be detected directly, and, especially in the radio band, they can have highly variable shapes, so the data are \textit{folded} at the pulsar's spin period to produce an average pulse profile which is instead quite stable and constitutes a recognizable ``fingerprint'' for a given pulsar.
Folding is done by ``cutting'' the time series in identical segments whose length is the pulsar's spin period. The segments are then averaged together, and if the segments are of exactly the right length, the signal-to-noise ratio of the resulting profile is maximized.
When the time series is constituted by photons, folding is done by doing a histogram of the phase of the photon, basically the fraction of the pulsar's spin period that has elapsed since the last pulse.

A new TOA is usually defined as the time of arrival of the peak (or another fiducial point) of the folded pulse profile, and that gives phase 0.
The maximum can be determined precisely by modeling the pulse profile with a series of Gaussian components or equivalent curves.
When calculating TOAs from a given instrument, if the pulse profile does not evolve with time, the procedure is generally done by calculating a cross-correlation with a template profile and/or using the FFTFIT algorithm that is a chi-square fit of a linear phase gradient between the Fourier representations of a template and the folded profile \citep{taylorPulsarTimingRelativistic1992}.
The template is usually a model of a high signal-to-noise profile of the same pulsar.
In high-energy astronomy, alternative likelihood-based approaches are sometimes used \citep[e.g.]{livingstoneXRAYRADIOTIMING2009,rayPreciseGrayTiming2011,clarkPSRJ1906+0722Elusive2015,niederDetectionTimingGammaRay2019}.
In this paper, we use a method based on maximum-likelihood fitting of pulse profiles (avoiding the \textit{a priori} definition of a template), see Section~\ref{sec:toa}.

\subsection{X-ray and Gamma-ray data processing}

Pulsar timing requires extreme attention to the detail of how the measurement times are recorded.

This paper uses data from 15 missions, each with its own specific procedure for data reduction, which usually follows the same pattern: using multiple sources of data, from ``housekeeping'' measurements recorded on the spacecraft to space weather information, mission-specific pipelines produce ``cleaned'' datasets in the form of ``event files'', which are typically FITS files with a table containing a time stamp and a number of physical properties for each detection that fulfils some quality criteria.
The details of each mission's pre-processing pipelines can be found in the Appendix~\ref{sec:missiondataproc}.

The time stamps of events are usually expressed in the mission elapsed time (MET), which give the number of seconds from a reference modified julian date (MJD), called MJDREF, typically using the Terrestrial Time scale (for definitions of time scales and conventions, see Appendix~\ref{sec:timedef}).
The MJDREF is written in the header of the FITS file, either as a single keyword, or as the combination of its integral part MJDREFI and the fractional part MJDREFF\footnote{\url{https://heasarc.gsfc.nasa.gov/docs/heasarc/ofwg/docs/summary/ogip_93_003_summary.html}}.
When this happens, sometimes the MJDREFF encodes the number of leap seconds (divided by 86400 to express them in days). In this case, using just MJDREFI can give times in the UTC scale, while summing MJDREFF + MJDREFI gives times in the TT scale.
Often, a TIMEZERO keyword adds a global correction, expressed in seconds, to the event times. In detectors where the measurement duration (hereafter ``frame time'') is significant, the TIMEDEL keyword is used to indicate the frame time, and TIMEPIXR to specify if the timestamps are referred to the start (0), the middle (0.5) or the end (1) of the frame time.
Finally, some missions carry time references whose alignment to UT drifts (e.g. quartz oscillators whose frequency can change depending on temperature), and for these missions an additional correction to their time reference can be provided on the ground.
We refer to this additional term as FINECLOCK below.
The total time in the TT scale, measured at the spacecraft, can then be calculated as\footnote{\url{https://heasarc.gsfc.nasa.gov/docs/xte/abc/time_tutorial.html}}
\texttt{MJD(TT) = (MJDREFI + MJDREFF)} + \texttt{(TIME + TIMEZERO + FINECLOCK + (0.5 - TIMEPIXR) * TIMEDEL) / 86400}.
Different time scales are allowed by the FITS standard, provided that the TIMESYS keyword is different from TT.
As mentioned before, in some missions MJD(UTC) can be calculated by just omitting the MJDREFF value in the formula above.

In radio pulsar studies, it is common to calculate the arrival time of pulsations through folding in the topocentric frame (so, as a UT time at the location of the radio telescope) using atomic clocks as a reference, and then to convert to the SSB during the modeling phase, depending on the version of the ephemeris used in the model.

In high-energy missions, the detection of pulses often requires integration over multiple orbits of the satellites, making topocentric TOAs impractical.
Therefore, in X-ray astronomy the conversion of times to the SSB, or \textit{barycentering}, often happens \textit{before} the folding, using tools like the \texttt{barycorr} FTOOL together with information on the satellite's position and velocity encoded in the orbit/attitude files shared by science operation centers as part of the ``housekeeping'' data for each observation.

The coordinates of the source are a key parameter for barycentering, as we are going to detail in Section~\ref{sec:consistency}. Different barycentering tools encode the coordinates used in different FITS keywords (e.g. \texttt{{RA,DEC}\_OBJ} for \texttt{barycorr}, \texttt{{RA,DEC}\_TDB} for HEASOFT's \texttt{barycen}).

All the relevant data are retrieved from the FITS files using Stingray \citep{huppenkothenStingrayModernPython2019}.

\subsection{Consistency in ephemerides, coordinates, reference frames}\label{sec:consistency}

Whether the barycentering is done before or after folding, the procedure involves a number of very delicate steps and decisions, which can introduce systematic errors in the TOAs.
The solar system barycenter is defined as the center of mass of the solar system, which is not at the center of the Sun, but rather at a point that moves around it due to the gravitational pull of the planets and the other objects in the solar system.
This is done through the use of solar system ephemerides, which are mathematical models that describe the positions and motions of the solar system bodies.
In X-ray astronomy it is customary to use the JPL development ephemerides, which are produced by the Jet Propulsion Laboratory and are the result of the numerical integration of the equations of motion of solar system objects\footnote{Alternative ephemerides exist, like the ones from the Institute of Celestial Mechanics and Ephemeris Calculation (IMCCE), called INPOP, and the ones from the Institute of Applied Astronomy, Russian Academy of Sciences (IAA RAS), called EPM, providing similar precision \citep{moiseevEphemerisTheoriesJPL2024}}

The JPL ephemerides are updated regularly and are indicated by a number, such as DE200, DE405, DE421, DE430, etc.
The choice of the ephemeris can have a significant impact on the barycentering procedure, introducing systematic errors in the TOAs by up to a few milliseconds, which is often a significant fraction of the pulsar's spin period.
Another issue is the choice of TDB or TCB, as mentioned in Appendix~\ref{sec:timedef}.

Then, the position of the pulsar in the sky is also important, as it determines the light travel time from the pulsar to the SSB.
The error on the arrival time at our instrument produced by an error $\epsilon$ on the right ascension is given by the following formula:
\begin{equation}
    \delta \mathrm{TOA} = r [\cos(\alpha + \epsilon) - \cos\alpha]\cos\delta \approx -r\epsilon\sin\alpha\cos\delta
\end{equation}
where $r$ is the Earth-Sun distance in light-seconds, $\delta$ is the declination and $\alpha$ is the angle traced by the SSB, the Earth and the pulsar.
In practice, the maximum TOA shift during one year of observations is
\begin{equation}
    \delta \mathrm{TOA} \approx -2r\epsilon\cos\delta = 4.8\,   \mathrm{ms} \cdot \left(\frac{\epsilon}{\mathrm{arcsec}}\right) \cos\delta
\end{equation}
The effect is macroscopic and it is often used to refine the pulsar's position in the sky with a precision comparable to high-resolution VLBI or optical observations.
Proper motion can also produce a significant effect on the TOAs and be measured through pulsar timing.
One has to pay attention to the choice of coordinates: the right ascension and declination are usually given in the J2000 equinox in the ICRS coordinate frame\footnote{More appropriately, ICRS is the name of the coordinate system, while the frame should be called ICRF; however, this rather confusing naming is used by barycorr and we decided to use the same for simplicity.}, but JPL ephemerides DE200-DE402 use the older FK5 reference frame, which differs by a fraction of an arcsecond from the ICRS frame and produces a shift up to $50\,\microsec$ in our processing.

As long as the data are consistently barycentered with a single ephemeris, time system, and source position, different choices usually lead to small changes in the analysis results. However, when data are collected and pre-processed by multiple teams using inconsistent choices, systematics can compromise analysis.

\subsection{Using the Jodrell Bank ephemeris} \label{sec:jbe}
The Jodrell Bank monthly ephemerides (hereafter JBE; \citealt{lyneTwentyThreeYearsCrab1993}) tracks the rotational history of the Crab pulsar over time, publishing a new timing solution roughly every month.
This priceless dataset, with a time span of almost four decades, is the ideal reference for the timing of the Crab pulsar.
Data are shared in two different formats: a space-separated text file with the spin frequency, one spin derivative and a TOA at the barycenter of the solar system, and a CGRO format\footnote{\url{https://www.jb.man.ac.uk/research/pulsar/crab/CGRO_format.html}; note that the coordinates quoted in this file correspond to (R.A., Dec.) 05:34:31.972 +22:00:52.070, 4.5 milliarcseconds from the official JBE coordinates 05:34:31.97232 +22:00:52.069, and that this solution does not quote a dispersion measure, so it is only useful for non-radio data} file with two spin derivatives and a TOA at the geocenter, defined as the peak of the radio profile at infinite frequency (i.e., corrected for interstellar dispersion).
Say our X-ray observation was performed at MJD 56000.
The corresponding line in the CGRO-format file gives the solution between MJDs 55987--56018, comprising of a spin frequency $\dot{\nu}$=29.7025600117591\,Hz, a spin derivative $\dot{\nu}=3.70689\times10^{-10}$\,Hz\,s$^{-1}$ , and a second derivative $\ddot{\nu}=1.95\times10^{-20}$\,Hz\,s$^{-2}$.
Moreover, this solution includes a TOA at the geocenter (MJD 56003.000000285).
These ingredients should be enough for a model describing the phase of the Crab during the month with an rms scatter of 0.6 milliperiods, also declared in this solution.
However, the JBE ephemerides publish the timing solution at the solar system barycenter using the DE200 FK5 reference frame, which is quite outdated, and not accounting for proper motion, so assuming that the pulsar's position is fixed in the sky to the J2000 position (R.A., Dec.) 05:34:31.97232 +22:00:52.0690 in the FK5 frame, or (R.A., Dec.) 05:34:31.970490 +22:00:52.087768 in the ICRS frame\footnote{Throughout the paper, we use the functions in \texttt{astropy.coordinates} to perform the coordinate conversion, e.g. defining \texttt{coordFK5 = SkyCoord(ra, dec, frame="fk5")} and then using \texttt{coordFK5.icrs} for the conversion to ICRS}.
This, despite the significant proper motion by $\sim13$\,milliarcsec/yr, which implies that the pulsar has moved by more than 0.5\arcsec from the position quoted in the JBE \citep{ngProperMotionCrab2006,kaplanPreciseProperMotion2008,gaiacollaborationVizieROnlineData2020,linRadioParallaxCrab2023}.
A parallax of $0.53 \pm 0.06$\,mas has also been measured \citep{linRadioParallaxCrab2023}, but its effect on the TOAs is negligible ($\lesssim1\,\microsec$) for the scope of this paper.
See Table~\ref{tab:coords} for a comparison of the Crab pulsar coordinates quoted by the JBE and \citet{gaiacollaborationVizieROnlineData2020} in different reference frames.

\begin{table*}
\centering
\caption{Crab pulsar coordinates from JBE and Gaia \citep{gaiacollaborationVizieROnlineData2020} expressed in different reference frames with precision truncated to three digits. The uncertainties are much smaller than the differences between the measurements, which are dominated by the systematic differences between the reference frames.}\label{tab:coords}
\begin{tabular}{l c c c c}
\hline
\hline
Reference frame & \multicolumn{2}{c}{Jodrell Bank (Epoch 1970.0\footnote{The Crab notes document does not state an epoch. A review by \citet{taylorCatalog558Pulsars1993} quotes the same position citing \citet{mcnamaraPositionalDeterminationNP1971}, ref. 156 in Table 1. In McNamara, the epoch is 1971.0 but the right ascension is different --05:31:31.428 (B1950) instead of 05:31:31.405. The position is assumed to be fixed in the JBE, in any case, so the epoch has no practical effect.}}) & \multicolumn{2}{c}{Gaia (Epoch 2016.0)} \\
& RA & Dec & RA & Dec \\
\hline
FK4 (B1950) & 05:31:31.428 & +21:58:54.402 & 05:31:31.405 &+21:58:54.466 \\
FK5 (J2000) & 05:34:31.972 & +22:00:52.069 & 05:34:31.949 &+22:00:52.135 \\
ICRS (J2000) & 05:34:31.970 & +22:00:52.088 & 05:34:31.947 &+22:00:52.154 \\
\hline
\end{tabular}

\end{table*}

The reason is consistency: users of the ephemerides do not need to worry about using different reference frames for datasets spanning different epochs (Lyne, priv. comm.).
The use of the JBE for X-ray timing studies, therefore, forces to barycenter the data using the same set of outdated ephemerides and position, losing flexibility when gathering data from all different missions (that might have been barycentered using different conventions).

\section{Supporting multiple JPL ephemerides at once} \label{sec:jpleph}
For this work, we developed a technique to compare TOAs taken with different missions, each using different solar system ephemerides, while keeping the JBE as the reference.
Let us consider the \suz observation of the Crab performed at MJD 56000, and say we want to calculate the arrival time of the X-ray peak compared to the radio peak.
We could in principle use the solution listed above to fold the data and calculate a TOA.
Let us say that for our scientific work we already have data for this observation, barycentered using the JPL-DE430 solar system ephemeris and the most recent position of the source considering proper motion.
As we have seen using these data with the Jodrell Bank ephemeris might lead to a timing error up to a few milliseconds, which is three orders of magnitude above the precision we desire for our cross-calibration purposes.

In order to avoid re-barycentering thousands of observations from 13 missions using JPL-DE200 and the FK5 position, we developed the following procedure:
\begin{enumerate}
    \item We \textit{simulate} a set of infinite-frequency TOAs \textit{at the geocenter} using the JB timing solution above in the DE200 ephemeris, with a small scatter of 0.1 \microsec;
    \item The difference between TOAs at the geocenter using different ephemerides is negligible: for example, for \xmm's orbit, the widest in our mission sample, it amounts to $\lesssim2\,\microsec/{\rm arcsec}$; hence, we assume these same TOAs remain the same using the DE430 ephemeris;
    \item We fit a new timing solution to these TOAs, using the DE430 ephemeris and the same ICRS position used for barycentering the X-ray data;
    \item We verify that the scatter of the TOAs with the new solution is comparable with the error bars of the simulated data, and therefore, also much smaller than the declared rms scatter of the JBE; we find that the two frequency derivatives are always adequate to the task in the $\sim$monthly timescale of the JBE.
\end{enumerate}
At this point, we have a new timing solution that is consistent with the JBE, but using the DE430 ephemeris.
This procedure can be implemented straightforwardly using the PINT \citep{luoPINTModernSoftware2021} software package, that supports a large set of timing models and JPL ephemerides.
The solution can be saved in a parameter file in PINT \citep{luoPINTModernSoftware2021} format, containing all the relevant information.
The geocenter arrival time from the JBE is encoded in the new model as TZRMJD, using \texttt{0} as observatory.
Special care must be put in the choice of coordinates relevant to a given ephemeris. DE200 uses the FK5 frame, whilst DE402 and later uses ICRS. The difference of arrival times using the wrong coordinate frames can be substantial, in the order of milliseconds.

\subsection{Folding}\label{sec:folding}
Once we have obtained a suitable timing solution for our observation, we can use it to fold the event arrival times and obtain a pulse profile.
The pulse phase is calculated with the usual Taylor expansion in Eq.~\ref{eq:phase}, using the timing model obtained in Section~\ref{sec:jpleph},
and the pulse profile is just a histogram of pulse phases, between 0 and 1, using $N$ phase bins.
The frequency and frequency derivatives are encoded in the timing solution.
At this point, the fundamental missing ingredient is $\phi_0$.
This has to be calculated so that the reference TOA in the JBE solution is at phase 0.
If we do this, since the phases are already referred to the radio TOA, our folded profiles will have automatically the phase 0 at the peak of the radio TOA.
All the relevant time conversions are done by PINT, by using the parameter file created with the procedure in Section~\ref{sec:jpleph}.
The phase of each photon can be calculated by properly converting the event arrival times from mission-elapsed seconds to MJD and using the \texttt{TimingModel.phase} method.
Observations spanning multiple JBE solutions (e.g. from Fermi, IXPE) were split in the relevant time ranges. In very high-throughput observations with more than 100M photons (e.g. frequently from NICER) we calculated a TOA every $\sim$50M photons.

\subsection{Deadtime correction of pulse profiles}\label{sec:deadtime}
\begin{figure}
    \centering
    \includegraphics[width=\columnwidth]{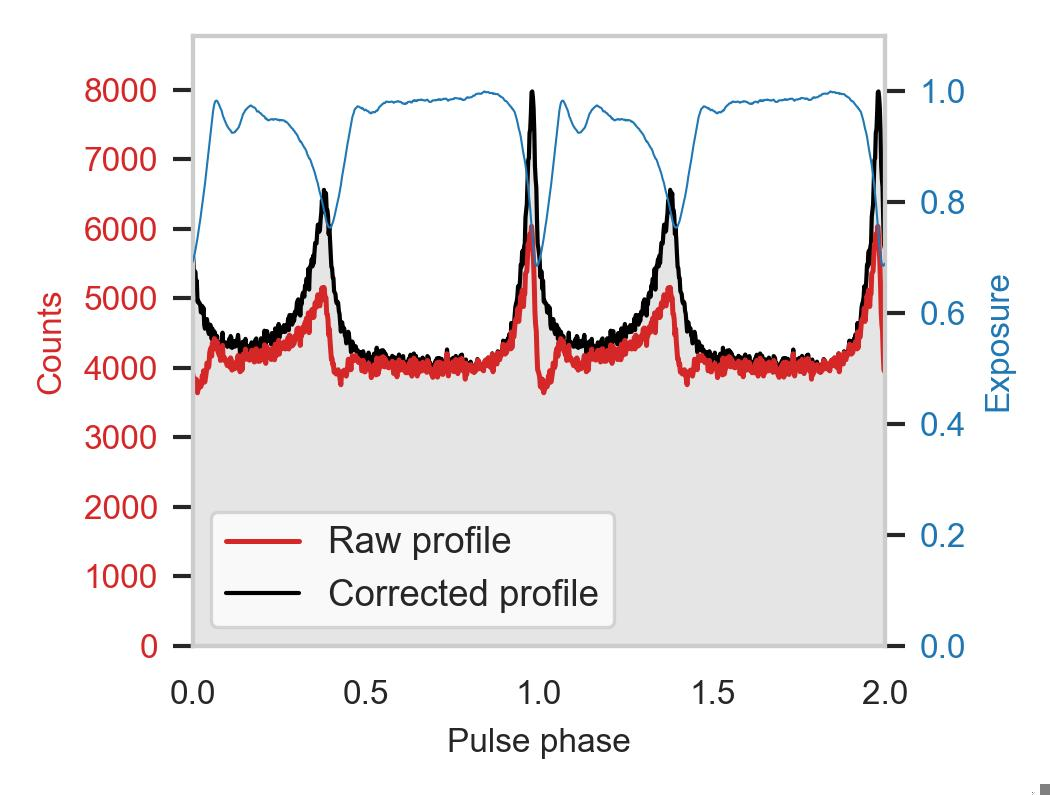}
    \caption{Example of deadtime correction of a \nustar pulse profile (ObsID 10002001009), using the method described in Section~\ref{sec:deadtime}. The blue line shows the exposure correction, the red line the raw profile and the black line the deadtime-corrected profile.}
    \label{fig:deadtime}
\end{figure}
For missions affected by dead time, like \nustar, we corrected the pulse profiles using the PRIOR column in the event list, which gives the live time before the current event, largely following the procedure used by \citet{madsenBroadbandXrayImaging2015}.
In short, the live time before each event was used to estimate the effective exposure in each bin of the pulse profile.
For example, let us consider a photon arrived at time $t_0$, with a value in the PRIOR column of 1\,ms.
This means that the live time before this photon started at $t_p=t_0-1\,\mathrm{ms}$.
Let us call $\phi_0$ the phase of the pulse profile corresponding to $t_0$, and $\phi_p$ the phase corresponding to $t_p$.
Our procedure will add 1 to all phase bins between $\phi_p$ and $\phi_0$, properly accounting for fractional bin coverage and phase wraps at the start and end of each pulse profile.
In the end, the exposure will be normalized to a maximum of 1, and the pulse profile divided by this effective exposure (so that more deadtime-affected pulse bins will be amplified).
This procedure is made available in the publically available code \texttt{pulse\_deadtime\_fix} \citep{matteo_bachetti_2025_16893714}.
The result of this procedure is shown in Figure~\ref{fig:deadtime}.

For \chandra/HRC data, affected by flux nonlinearity at high count rate that are not well described by a dead time model, we apply a correction factor
\begin{equation}
r_{\rm corr} = 45.91 - \sqrt{2058.4 - 75.6\,\, r_{\rm det}}
\end{equation}
between 12 and 24 counts per second to each bin of the folded profile, following \citet{peaseCountRateLinearity1998}.

\subsection{Residual calculation} \label{sec:toa}
\begin{figure}
    \centering
    \includegraphics[width=\columnwidth]{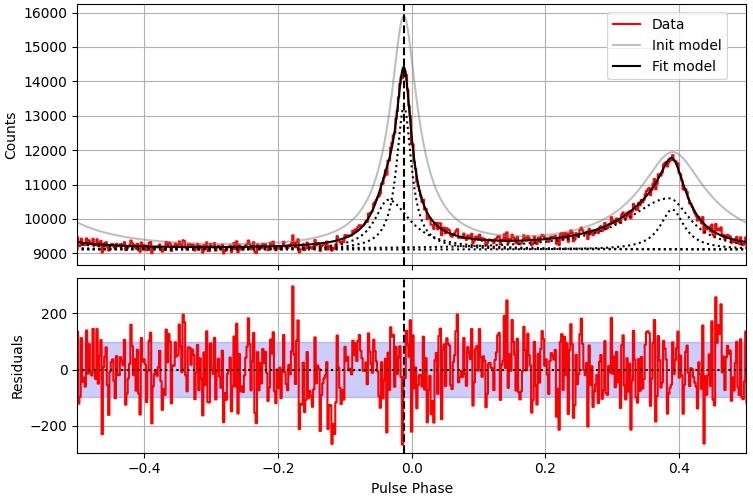}
    \caption{Folded profile from \nicer (obsid 101301013) modeled with the combination of symmetric and asymmetric Lorentzians described in Section~\ref{sec:toa}.}
    \label{fig:profile_fit}
\end{figure}
Once we have the folded profile, as we saw in Section~\ref{sec:folding}, the phase 0 of the profile should coincide with the main radio peak.
As is very well known from the literature \citep{rotsAbsoluteTimingCrab2004}, the peak in the X-rays will not coincide with the radio peak, and we need a way to measure this delay, that can then be used to compare the performance of different high-energy missions.

There are mainly two methods used in the literature to calculate the phase of a pulsar: the alignment of the folded profile with a template (e.g. FFTFIT, \citealt{taylorPulsarTimingRelativistic1992}) or fitting a model to the pulse profile \citep[e.g.][]{nelsonOpticalTimingPulsar1970}.
The first method is the most stable in ideal conditions, when the template for the pulse is correct, bringing precise measurements even in noisy data.
However, if the template is not correct, as can happen if the response of the instrument used to calculate the template is different from the instrument used to obtain the folded data, or worse, if the folding is not perfect due to issues with the timing systems, this method can give unrealistically small errors and/or align to the wrong peaks.
In the case of the Crab pulsar, where low signal-to-noise is seldom a problem, the second method yields meaningful results both in ``good'' datasets without requiring an exact template, and in datasets affected by instrumental issues, where the template adapts to an imperfect profile measuring the peak position and providing a reasonable error bar.

For this work, we chose to model each peak as a combination of a symmetric Lorentzian curve plus an asymmetric Lorentzian curve.
The phases of the two Lorentzians composing each peak are linked (not more than a difference of 0.02 in phase).
All other parameters are left free to vary.
The folded profile changes considerably at different energies, until the secondary peak becomes comparable to the first peak making the identification less obvious.
However, the secondary peak maintains a distance of $\sim0.4$ in phase from the primary, so that the two peaks can easily be identified by their phase distance.
The error on the arrival time can be calculated from the error of the centroid of the sharpest Lorentzian.
We verified that the scatter of TOA residuals is indeed comparable to the error bars estimated in this way.
The fit quality can be seen in Figure~\ref{fig:profile_fit}.

\subsection{Calculation of local pulsar frequency}
We also calculated the local frequency of the pulsar by applying the quasi-fast folding algorithm \citep{bachettiAllOnceTransient2020}, as implemented in Stingray \citep{huppenkothenStingrayModernPython2019}, to each observation, searching a small frequency-frequency derivative plane around the nearest radio solution.
We can use this information in special cases near a glitch or when the JBE is missing, and we plan to exploit it in future versions for the determination of X-ray-only ephemerides of the Crab untied to the JBE.

\subsection{Validation}\label{sec:validation}
\begin{figure}
    \centering
    \includegraphics[width=\columnwidth]{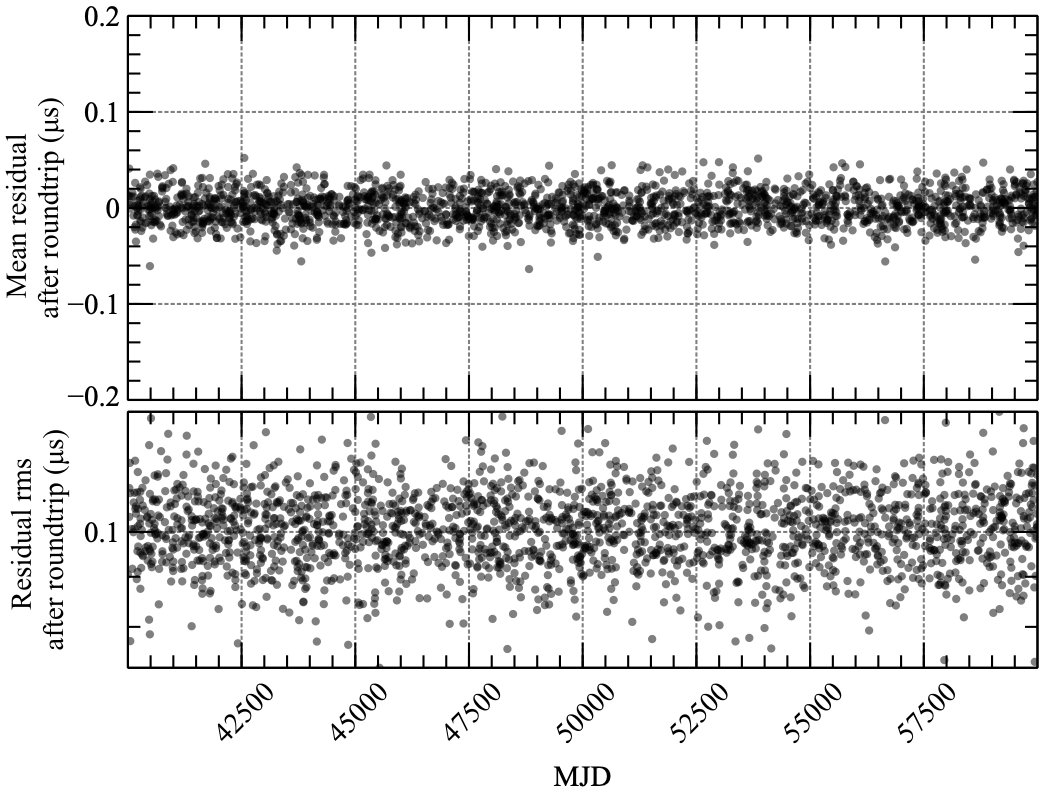}
    \caption{Results of the validation procedure for the model refitting described in Section~\ref{sec:validation}. We fit a DE430 model to DE200-simulated TOAs, then simulated TOAs with the new model and calculated the residuals to the original DE200 model. For each instance of the simulation, we plot the mean residual in the upper panel and the rms of the residuals in the lower panel. The procedure works remarkably well, with residuals and rms comparable or lower to the injected error bars in the simulated data (0.1\,\microsec).}
    \label{fig:validation}
\end{figure}

\begin{figure}
    \centering
    \includegraphics[width=\linewidth]{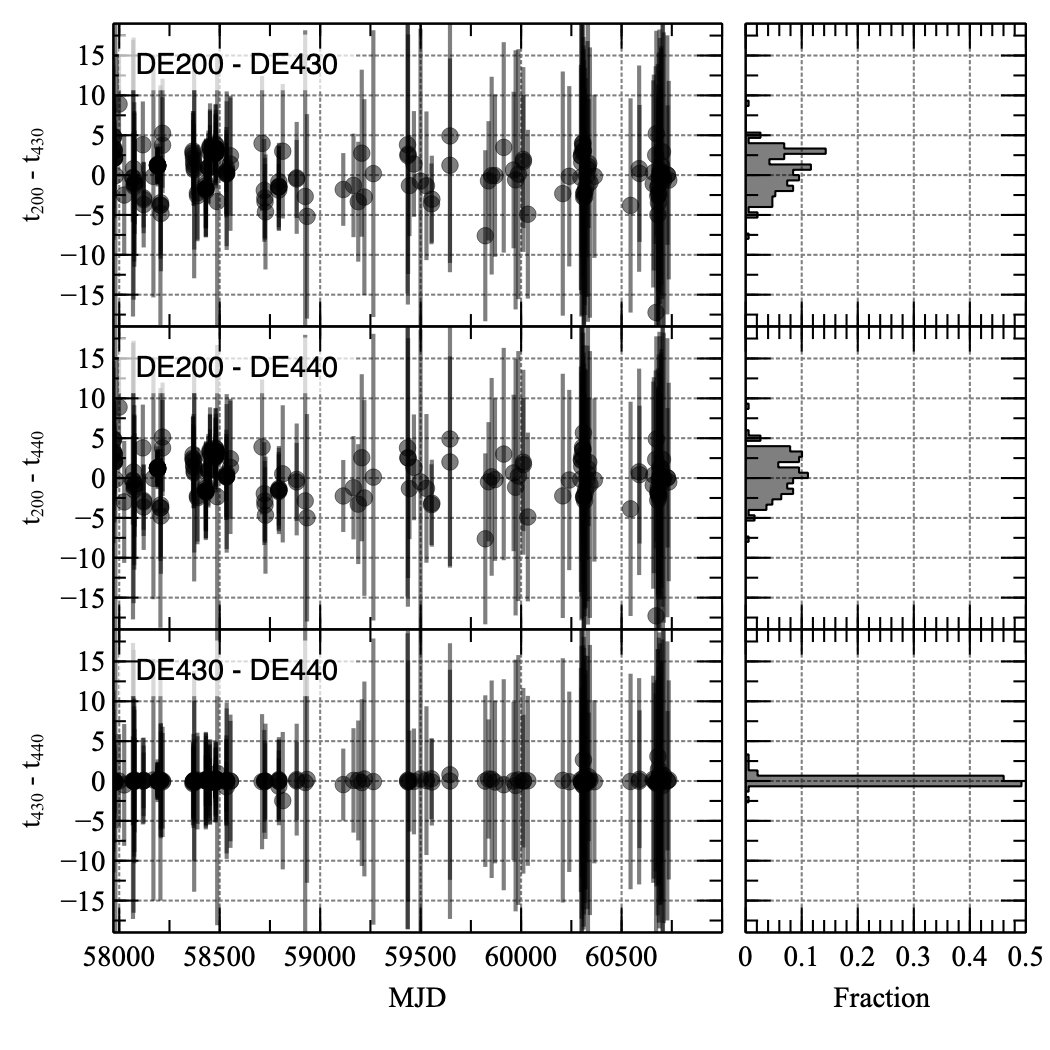}
    \caption{End-to-end validation of the procedure in Section~\ref{sec:jpleph} using \nicer datasets. (Left) Difference between the TOA residuals (labeled $\rm t_{eph}$) calculated from datasets barycentered using the DE200 ($\rm t_{200}$), DE430 ($\rm t_{430}$), and DE440 ($\rm t_{440}$) JPL ephemerides. (Left) a histogram of the corresponding cases.
    We did not consider observations where the TOA uncertainty was $>15\,\microsec$.
    The uncertainty of these TOAs was typically $\sim5\,\microsec$, and residual differences are always well inside error bars.}
    \label{fig:TOA_comparison}
\end{figure}
To validate the method described above, we made a number of tests, described in the following paragraphs.

\paragraph{Roundtrip of simulated TOAs.} Starting from a model of the Crab pulsar defined in the DE200 ephemeris, we simulated one month of geocentered TOAs as in Section~\ref{sec:jpleph}, we fit a DE430 model to them using a random position in the ICRS frame around the JBE position, simulated geocentered TOAs with the new model, and calculated the residuals to the initial model.
Results are shown in Figure~\ref{fig:validation}.
We found that, for the method to work with a precision of better than 5\,\microsec, quadruple precision\footnote{Here we refer to the data type often called \texttt{float128}, despite being in practice an 80-bit floating point in common architectures and, therefore, not actual quadruple precision.} floating points have to be available in the computer doing the analysis.
For example, in recent ARM-based Mac machines, the scatter in Figure~\ref{fig:validation} increased by a factor $\sim$20, with strong patterns dependent on the pulse frequency and the day of year.
The problem did not appear on Intel-based Macs, x86 Linux-powered machines, or when using the Rosetta emulator on ARM-based Macs (which is also capable of emulating long doubles).
Also, we found that the fit with the two-derivative model was always adequate, yielding an rms scatter of residual comparable with the injected 0.1\,\microsec errorbars, if the simulated position was $<5$\arcsec, which is typically the case in our observations, and still $\lesssim$1\,\microsec for a position error $<1$\arcmin.
\paragraph{End-to-end validation} We proceeded to an end-to-end validation the procedure using \nicer observations of the Crab pulsar, barycentered using the DE200, DE430, and DE440 ephemerides and folded using the timing models obtained with our method.
We folded the profiles and applied the residual calculation described in Section~\ref{sec:toa}, and then compared the results.
We find that the difference between the residuals using the DE200 and DE4xx ephemerides was always consistent with 0 (\figref{fig:TOA_comparison}).
This demonstrates that the method described in this paper is able to calculate TOAs from a variety of JPL ephemerides and positions, and that the differences are comparable to the TOA error at this level, with a possible systematic shift by $\sim2\,\rm\mu s$ that we are investigating.
This level of precision is adequate for Crab studies, but can be improved by analyzing sharper pulsars.
This is planned for future studies.

\subsection{Pipeline}
All steps above are automated in the \texttt{TOAExtractor} code\footnote{\url{https://matteobachetti.github.io/TOAextractor/}} v. 0.6.0 \citep{matteo_bachetti_2025_17634233}.
This code is structured as a pipeline, using the \texttt{luigi} orchestrator\footnote{\url{https://luigi.readthedocs.io/en/stable/}} v. 3.6.0.
This software allows for a simple and modular way to define the steps of the pipeline, and to run them in parallel over multiple files by using multiple ``workers''.
The end of each step of the pipeline is marked by the production of a file.
The appearance of this file triggers the start of the next step.
Failing steps do not produce the output file, blocking the pipeline on a given file. Fixing the error and re-running the pipeline will restart the processing from the failed step.
To reprocess a given dataset from one particular step of the pipeline, it is sufficient to delete all files produced by the step to be repeated and all subsequent steps, and re-run the pipeline.

\subsection{Plotting and quality assurance}

\begin{figure*}
    \centering
    \includegraphics[width=\linewidth]{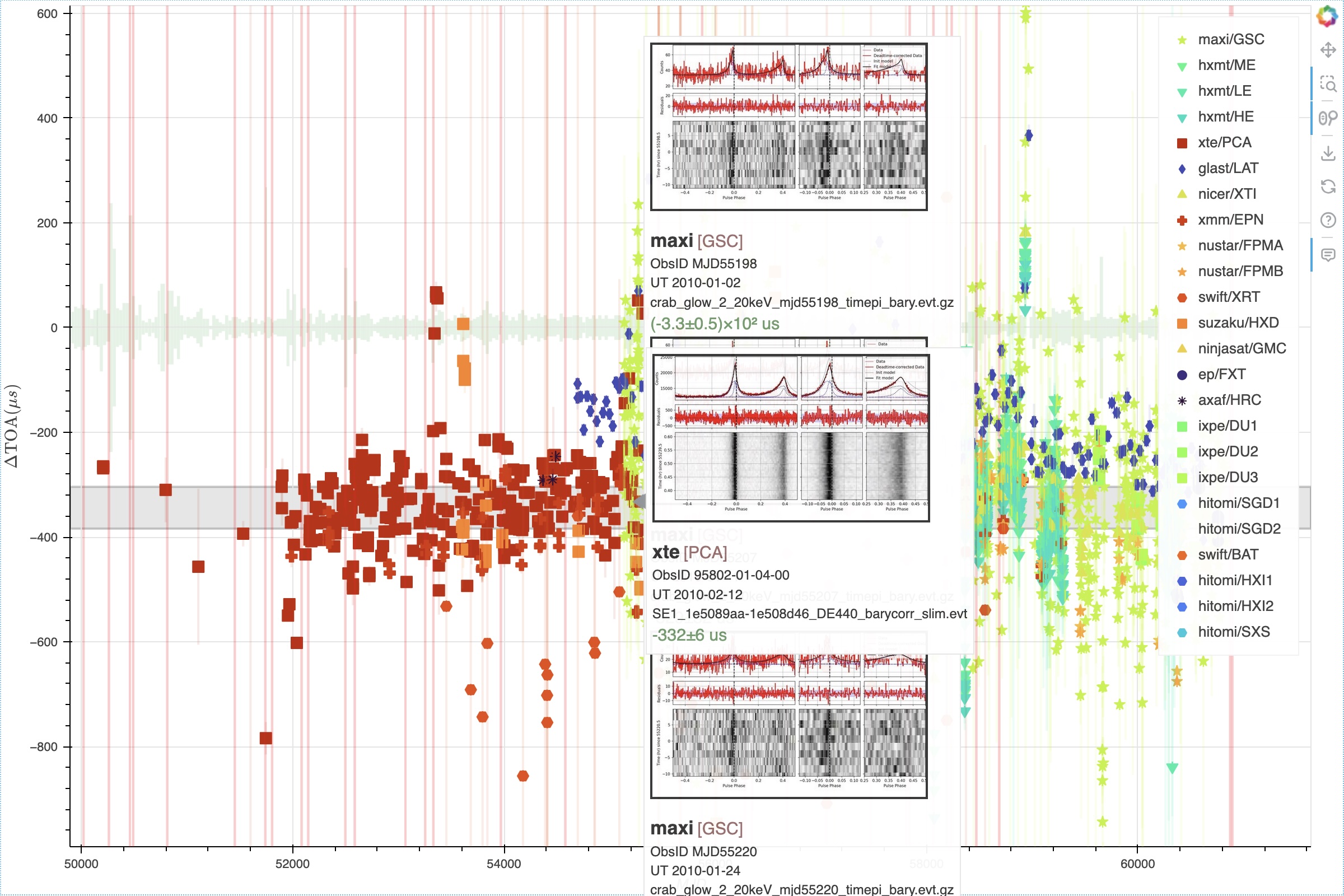}
    \caption{Interactive summary plot produced by \texttt{TOAExtractor}.
    This plot allows for a simple exploration of the residuals, by showing diagnostic information as the user hovers over each point. The horizontal grey band indicates the X-ray residual measured by \citet{rotsAbsoluteTimingCrab2004}.
    Vertical red lines indicate glitch epochs (In this static image it might be difficult to distinguish them from some large error bars from the data points, but the difference becomes clear in the interactive interface).
    Problematic observations and time intervals are easily identified: some observations with large residuals might be due to low count rates, temporary instrumental issues, or local issues with the radio ephemerides (e.g. the large deviation around April 2020, MJDs 58925--58975, due to the closure of the observatory for the COVID pandemic; recent glitches can also make the solution less reliable). }
    \label{fig:residuals}
\end{figure*}
This pipeline is conceived to be run on a large number of observations from different missions, possibly ever increasing over time.
As such, it is important to have a quick way to check the quality of the data.
For this reason, we developed an interactive summary plot (Figure~\ref{fig:residuals}) that allows a view on the results, helping to single out outliers and/or time intervals where the radio ephemeris was inadequate.
Each data point shows the delay of the X-ray peak with respect to the radio peak for a single observation.
Data from a given instrument can be shown or hidden by clicking on the legend.
By hovering over each point, the user can see the observation ID, the mission, the TOA, the error on the TOA, the energy range, and a diagnostic plot showing if the fit was successful.
Time intervals where the radio solution was inadequate (e.g. due to glitches) or absent (e.g., April 2020 due to the COVID pandemic) can be easily identified by the presence of a relatively large number of outliers from different missions.

\section{Results}\label{sec:results}
\begin{table*}[htb]
\centering
\caption{Summary of the results in this work, separated  by mission and instrument. The mean residuals $r_{\rm mean}$ are calculated as the median (if $N>$ 20) or the average of the residuals of the X-ray TOAs with respect to the radio TOAs; the standard deviation $\sigma$ similarly uses the median absolute deviation and the standard deviation in the two cases. For missions with too few observations we do not report the standard deviation. 
$\sigma_{\rm stat}$ is the median/mean statistical error on the single TOA measurement. \label{tab:summary}. Some long observations, or observations from high-throughput instruments, had sufficient counts to allow multiple TOAs. Missions with $>$1000 observations are not completely analyzed, the table shows the status at 2025-09-02}
\begin{tabular}{llcccccccc}    
Mission & Instrument & Min MJD & Max MJD & $N$ & $r_{\rm mean}$
& $\sigma$
& $\sigma_{\rm stat}$
& Rate
& Pulsed fraction\\
& & & & & \microsec
& \microsec
& \microsec
& Counts s$^{-1}$
& \% \\
\hline
\hline
AXAF/Chandra/CXO & HRC & 51574.28 & 56589.14 & 11 & -458 & 310 & 83 & 2.42 & 100 \\
EP & FXT & 60368.38 & 60758.73 & 7 & -391 & 79 & 15 & 1490 & 34 \\
GLAST/Fermi & LAT & 54696.08 & 60617.92 & 196 & -167 & 80 & 14 & 0.00498 & 84 \\
HITOMI & HXI1 & 57472.63 & 57472.75 & 2 & -264 & -- & 27 & 73.1 & 61 \\
HITOMI & HXI2 & 57472.63 & 57472.74 & 2 & -214 & -- & 63 & 68.6 & 57 \\
HITOMI & SGD1 & 57472.63 & 57472.63 & 1 & -244 & -- & 78 & 1.17 & 52 \\
HITOMI & SGD2 & 57472.63 & 57472.63 & 1 & -157 & -- & 690 & 0.217 & 50 \\
HITOMI & SXS & 57472.64 & 57472.64 & 1 & -106 & -- & 16 & 126 & 46 \\
HXMT & HE & 57992.43 & 60398.80 & 342 & -370 & 90 & 12 & 1360 & 36 \\
HXMT & LE & 57992.60 & 60398.96 & 385 & -319$^a$ & 100 & 61 & 1220 & 38 \\
HXMT & ME & 57992.26 & 60398.95 & 518 & -332 & 86 & 18 & 439 & 37 \\
IXPE & DU1 & 59632.22 & 60727.75 & 7 & -350 & 81 & 7.6 & 2.81 & 77 \\
IXPE & DU2 & 59632.22 & 60727.75 & 7 & -352 & 76 & 8.5 & 2.5 & 75 \\
IXPE & DU3 & 59632.22 & 60727.75 & 7 & -345 & 80 & 7.7 & 2.45 & 76 \\
MAXI & GSC & 55135.51 & 60866.50 & 3847 & -312 & 200 & 120 & 0.42 & 49 \\
NICER & XTI & 57970.82 & 60364.52 & 145 & -314 & 90 & 7 & 10400 & 38 \\
NINJASAT & GMC & 60365.09 & 60771.04 & 14 & -406 & 53 & 22 & 26 & 44 \\
NUSTAR & FPMA & 56132.26 & 60388.08 & 79 & -328 & 79 & 12 & 417 & 50 \\
NUSTAR & FPMB & 56132.26 & 60388.08 & 78 & -329 & 76 & 14 & 394 & 49 \\
SUZAKU & HXD & 53604.23 & 56722.55 & 18 & -326 & 170 & 310 & 53 & 39 \\
SWIFT & BAT & 57431.04 & 58190.06 & 3 & 20.5 & 160 & 58 & 6770 & 6.7 \\
SWIFT & XRT & 53454.21 & 58725.60 & 43 & -615 & 150 & 1200 & 644 & 21 \\
XMM & EPN & 51632.96 & 59269.92 & 97 & -356 & 90 & 15 & 178 & 42 \\
XRISM & RESOLVE & 60388.50 & 60388.50 & 1 & -312 & -- & 15 & 38.3 & 36 \\
XTE & PCA & 50205.36 & 55927.00 & 312 & -333 & 74 & 6.1 & 10600 & 46 \\
\hline
\multicolumn{10}{l}{(a) We applied a known offset of -864 ${\rm \mu s}$ to the measured value \citep{tuoInorbitTimingCalibration2022}}
\end{tabular}
\end{table*}

For the purposes of this paper, we will focus on the results that are relevant to the cross-calibration efforts of IACHEC. We will leave the scientific analysis of the Crab pulsar to a future paper, and we will release the database of TOAs obtained in this work to the general public to encourage the scientific interpretation of results.
The code used to produce the TOAs is also open source and available for general use.

We applied the method described above to thousands of observations of the Crab pulsar involving 15 missions, spanning 29 years of observations, from 1996 to 2025.

\begin{figure}
    \centering
    \includegraphics[width=\linewidth]{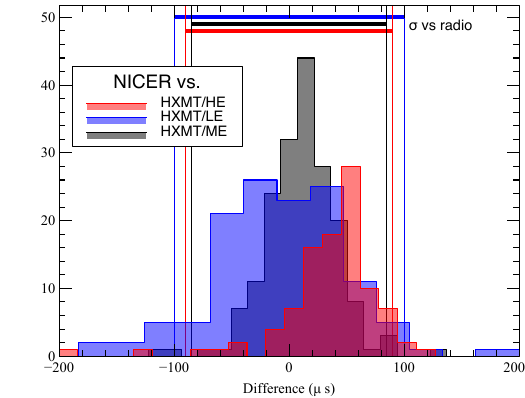}
    \caption{Difference between NICER/XTI and the three instruments onboard HXMT, at times where a TOA was calculated with both missions at a distance of less than one day. The shift of the X-ray profile at different energies causes the centroid of the distributions to increasingly depart from zero as the energy increases. We plot horizontal lines showing the -- significantly larger -- scatter obtained when using the radio solution as reference (from Table~\ref{tab:summary}). The delays for HXMT/LE are adjusted by 864$\rm \mu s$ as in Table~\ref{tab:summary} for graphical purposes, and their positioning around 0 mostly confirms the results of \citealt{tuoInorbitTimingCalibration2022}.}
    \label{fig:nicervshxmt}
\end{figure}
This wealth of data allows us to measure the relative timing accuracy of different missions, and to compare the results with the JBE (Figure~\ref{fig:nicervshxmt}).
The X-ray TOAs are typically leading the radio TOAs, as reported in the past \citep[e.g.][]{rotsAbsoluteTimingCrab2004,smithPulsarTimingFermi2008,cusumanoTimingAccuracySwift2012}.
The time difference between the X-ray and radio TOAs is not consistent with being constant, but it varies with the energy range of the photons, and notably, with time.

The variation of the X-ray/radio TOA difference with time is often larger than the statistical error on the TOAs, but the TOAs from the different missions are consistent with each other within each mission's requirement.
This might be caused by a number of factors, which we briefly discuss below.

\paragraph{Systematic errors in the JBE.} The JBE in CGRO format typically declares a rms scatter of the TOAs ranging from 0.5 to 2 milliperiods (i.e. $\sim$15--70 \microsec), depending on the time interval and the number of observations used to calculate the timing solution. This appears to be the largest source of uncertainty in the TOA, and is able to explain most of the scatter in our residuals.

\paragraph{Systematic errors in the method.} We investigated the possibility that the method described in Section~\ref{sec:jpleph} introduces systematic errors in the TOAs.
We verified that the scatter of the TOAs after the roundtrip procedure is typically of the order of the uncertainties of the simulated data (Figure~\ref{fig:validation}).
Moreover, we repeated the procedure using data barycentered with different JPL ephemerides and source positions, including the exact JBE position, and we found that the TOAs are consistent within 5\,${\mu}$s or the uncertainties of the TOA calculation (Figure~\ref{fig:TOA_comparison}).

\paragraph{Intrinsic variability of the X-ray/radio delay.} In a few, selected, intervals, the X-ray/radio delay appears to be larger than the statistical error on the TOAs or the expected systematic errors, and to vary with time consistently in X-ray data, but departing from a constant radio-X delay. More work is needed to investigate this intriguing possibility, using the software developed in this work to analyze even larger datasets.

\section{Conclusions}\label{sec:conclusions}

In this work, we presented a method to compare pulsar times of arrival (TOAs) from different space missions, even when using heterogeneous timing solutions, JPL ephemerides, and source positions.
The method is based on the simulation of geocentered TOAs using a reference timing solution, and then fitting a new timing solution to these TOAs using the ephemerides and source position used for barycentering the data.
We validated the method by simulating and comparing TOAs across different ephemerides (DE200--DE440), and verified that the TOAs are consistent within 5 \microsec or the uncertainties of the TOA calculation.
We applied the method to thousands of Crab pulsar observations from 15 missions over 29 years (1996--2025), enabling robust cross-calibration of timing accuracy.
We find good agreement between the TOAs from GPS-carrying missions like \nicer/XTI and \hxmt/ME, with mean offsets within a few microseconds relative to each other, and scatter consistent with the expected statistical and systematic uncertainties.
This demonstrates the robustness of the method for high-precision timing cross-calibration across modern missions.
We released open-source tools and a TOA database to support future calibration and scientific studies using the Crab pulsar, and plan to extend the software and database to include additional missions and pulsars.

\begin{acknowledgments}
The authors wish to thank the anonymous referee for their insightful and constructive comments, which contributed greatly to the clarity and soundness of the paper.
Thanks also to Andrew Lyne, Paul Ray, George Younes, David Smith, Allyn Tennant for thoughtful discussions.
MB wishes to acknowledge partial support by Italian Research Center on High Performance Computing Big Data and Quantum Computing (ICSC), project funded by European Union - NextGenerationEU - and National Recovery and Resilience Plan (NRRP) - Mission 4 Component 2 within the activities of Spoke 3 (Astrophysics and Cosmos Observations).
This work was supported by the JSPS Core-to-Core Program (grant number: JPJSCCA20220002) and Japan Society for the Promotion of Science Grants-in-Aid for Scientific Research (KAKENHI) Grant Number JP20K04009 (YT), JP19K14762, JP23K03459, JP24H01812 (MS).
This work uses data obtained with Einstein Probe, a space mission supported by the Strategic Priority Program on Space Science of the Chinese Academy of Sciences, in collaboration with ESA, MPE, and CNES (Grant No. XDA15310303, No. XDA15310103, No. XDA15052100).
VLK and AR acknowledge support from NASA contract to the Chandra X-ray Center NAS8-03060.
This work made significant use of the Fornax Science Console, which is part of the NASA Astrophysics cloud-based Fornax Initiative jointly developed by Goddard Space Flight Center’s Astrophysics Projects Division (GSFC/ApPD) and the astrophysics archives – the High Energy Astrophysics Science Archive Research Center (HEASARC), the Infrared Science Archive (IRSA), and the Mikulski Archive for Space Telescopes (MAST).

\end{acknowledgments}

\vspace{5mm}
\facilities{CXO, EP, Fermi, Hitomi, HXMT, IXPE, MAXI, NICER, Ninjasat, NuSTAR, Suzaku, Swift (XRT and BAT), XMM, XRISM, XTE.}

\software{astropy \citep{astropycollaborationAstropyCommunityPython2013,astropycollaborationAstropyProjectBuilding2018a,astropycollaborationAstropyProjectSustaining2022},
          Stingray \citep{huppenkothenStingrayModernPython2019},
          PINT \citep{luoPINTModernSoftware2021},
          TOA extractor \citep{matteo_bachetti_2025_17634233}
          }



\appendix

\section{Time definitions and conventions}\label{sec:timedef}

\begin{figure*}
    \centering
    \includegraphics[width=\linewidth]{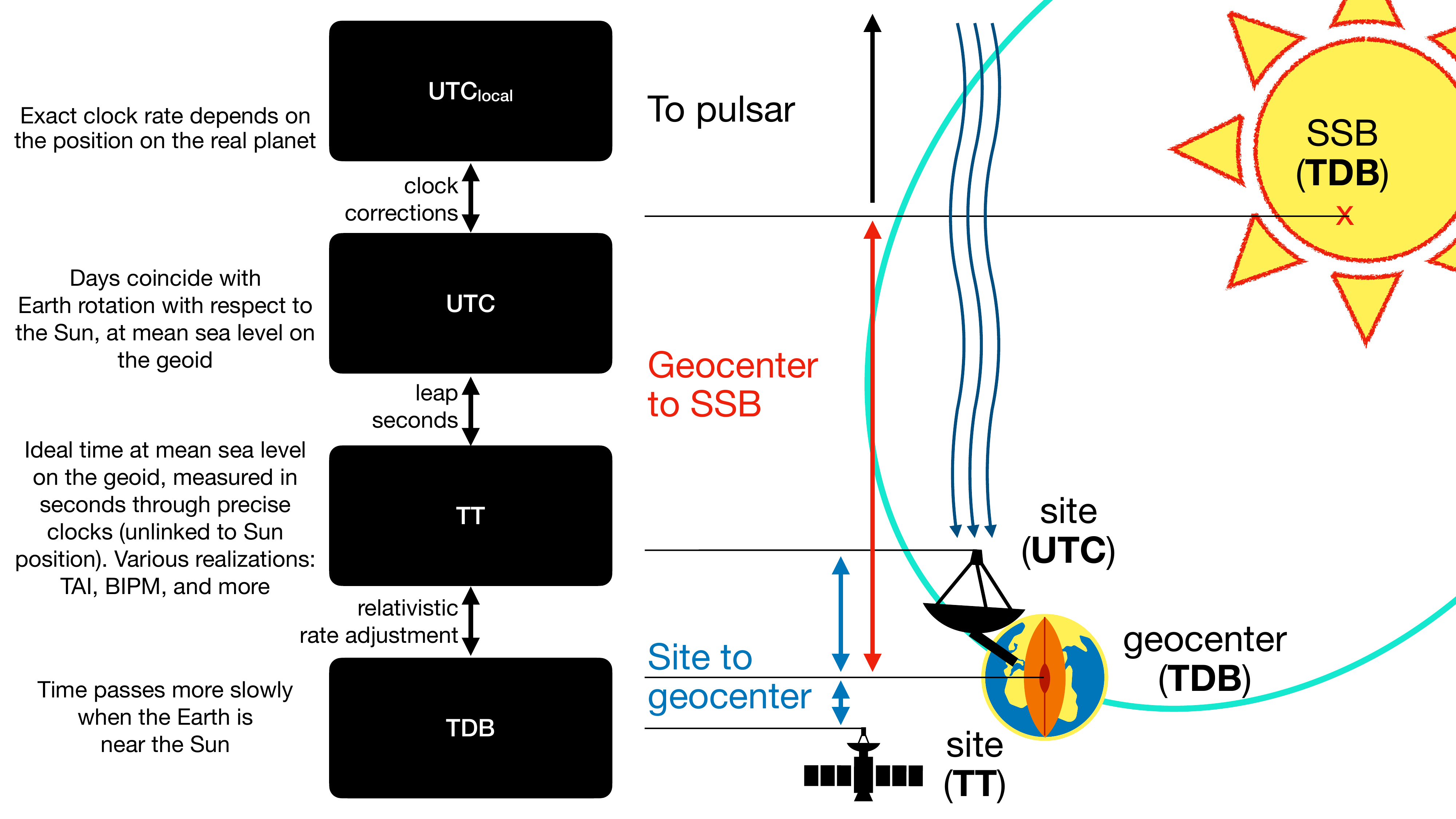}
    \caption{Relevant time scales and transformations described in Section~\ref{sec:timedef}.}
    \label{fig:timescales}
\end{figure*}

When doing precise timing work with pulsars, it is important to agree on two fundamental aspects: the time scale (which decides how fast our clocks run and a specific reference time to refer our time measurements to), and the relative position of the observer, the pulsar, and the solar system barycenter (which allows us to correct for different light travel times from the source to our instrument as the Earth moves around its orbit).
What we summarize here is treated in great detail in the TEMPO2 paper II by \citet{edwardsTEMPO2NewPulsar2006a}.
There is also some confusion on the acronyms used by technical reports on time scales and in pulsar astronomy papers, that we try to address here.

\subsection{Time scales}
The calendar utilized in everyday tasks is based on the duration of a day, which is defined as a rotation of the Earth with respect to the Sun\footnote{\textit{not} a full rotation on its axis, which would be a \textit{sidereal} day}, and initially second was defined as 1/86400th of the duration of an average day.
However, this process does not last a constant amount of time, but the Earth's rotation is not stable and, over time, tends to slows down, mostly due to the tidal effect of the Moon.
Since 1967, the second has become an International System of Units (SI) base unit, defined as the duration of 9,192,631,770 periods of the radiation corresponding to the transition between the two hyperfine levels of the ground state of the $^{133}$Cs atom \citep{bureauinternationaldespoidsetmesuresInternationalSystemUnits2025}.
A day is now defined as 86400\,s, and to correct for the Earth rotation evolution, leap seconds are introduced from time to time to synchronize the calendar to the rotation (so that there are single days of 86401 seconds, like December 31, 2016).
The time scale tied to the Earth rotation is called \textbf{Coordinated Universal Time (UTC)}.
It is calculated as a byproduct of the \textbf{Terrestrial Time} (TT), which is a theoretical time measured on the surface of a spherical Earth whose radius is the mean sea level.
The most commonly used estimate (or \textit{realization}) of the TT scale is the \textbf{international atomic time} (TAI), calculated from the weighted (and corrected for relativistic effects) average of 450 atomic clocks, mostly caesium clocks. The Bureau International de Poids et Measures (BIPM) annually reanalizes the same clock time series used by TAI and produces an improved TT solution that can be used for the most challenging metrology applications.
Independent realization of TT are under study, notably one based on pulsars \citep{hobbsPulsarbasedTimescaleInternational2020}, which would definitively disentangle our measurement of time from the Earth's influence and be suited to interstellar travel.

Because of general relativity, the pace at which clocks run is a function of their acceleration (or, equivalently, the gravitational field they are subject to).
A clock at rest in empty space will run faster than one close, e.g., to Earth; different clocks at different positions on the geoid will run at different paces, being subject to a slightly different gravitational pull; and as the Earth moves around the Sun in its elliptical orbit, the rate at which the ideal clock from TT runs remains non constant, due to the distance from the Sun. This effect is called Einstein delay \citep{damourGeneralRelativisticCelestial1986}. This gives rise to a periodic change of clock rates which needs to be taken into account in precision timing. The time scale correcting TT for these relativistic rate changes is called \textbf{barycentric dynamical time} (TDB).
An even more fundamental scale is the barycentric coordinate time (TCB), which is equivalent to the proper time measured by a clock which is co-moving with the barycenter of the Solar system but outside its gravitational well, calculated in a fully relativistic framework. In practice, TCB differs from TDB by a fixed rate\footnote{IAU 2006 Resolution 3: \url{https://syrte.obspm.fr/iauJD16/IAU2006_Resol3.pdf}}, and TDB is nowadays defined in terms of TCB.
The Tempo2 package \citep{hobbsTEMPO2NewPulsar2006} internally uses TCB, while TEMPO \citep{ASCLnetTempoPulsar} and PINT \citep{luoPINTModernSoftware2021} use TDB. TDB is also still used as an independent variable in the JPL-DE ephemerides.
In parallel, the Solar System Einstein delay is computed using the IF99 model \citep{irwinNumericalTimeEphemeris1999} by Tempo2 and the FB90 \citep{fairheadAnalyticalFormulaTime1990} by TEMPO and PINT\footnote{They are specified through the TIMEEPH parameter in parameter files}.

As a side note, despite the name similarity, TDB is a time scale valid everywhere in the relativistic framework. It does not necessarily involve the so-called \textit{barycentering} of the data, which consists of correcting the time measurements for light travel time across the solar system (see below).

\subsection{Observer position-related transformations}
Once we have agreed on a time scale and our clocks are synchronized, we still need to deal with the fact that signals travel in space at a finite velocity, the speed of light.

Depending on the position of the Earth in its orbit, the arrival time at the instrument of a given photon can change by up to 16 minutes (Roemer delay).
The same applies to smaller but significant corrections by milliseconds depending on the instrument position on the Earth surface or in orbit.
It is therefore important to find a way to express the time of arrival at a fixed reference point, typically the solar system barycenter (SSB).
Correcting for this effect is called \textit{barycentering}.

Modern pulsar timing codes use the detailed models of the solar system provided by the JPL-DE in order to determine not only the position of the SSB, but also other kinds of delay due to the signal traversing the curved space time near the Sun and other Solar System objects, the so-called Shapiro Delay \citep{shapiroFourthTestGeneral1964,damourGeneralRelativisticCelestial1986}.

\section{Mission-specific data preparation}\label{sec:missiondataproc}
In this Appendix, we describe the data preparation steps specific to each mission. Some apparent inconsistency might arise from differences in the data processing pipelines and calibration procedures employed by each mission. In all cases, the most important step for our work is the barycentering procedure. Wherever not specified, we assume that DE200 ephemerides use the FK5 reference frame, and ICRS otherwise.

\subsection{Chandra/HRC}
\begin{table}[]
    \centering
    \caption{{\sl Chandra}/HRC observations of Crab$^a$}
    \label{tab:chandraHRC}
    \begin{tabular}{l c c c}
    \hline\hline
    \multicolumn{2}{l}{Observation} & Exposure & RA,Dec \\
    ID & Start Time & [ks] & ICRS 2000 \\
    \hline
758 & 2000-01-31 01:21:19 & 74.42 & 5:34:31.9242,+22:00:51.920 \\
8548 & 2007-09-26 11:51:37 & 7.71 & 5:34:31.9471,+22:00:52.103 \\
9764 & 2007-12-27 11:39:14 & 3.62 & 5:34:31.9349,+22:00:51.841 \\
9765 & 2008-01-22 16:04:52 & 97.28 & 5:34:31.9317,+22:00:51.937 \\
11245 & 2010-11-16 07:11:01 & 19.96 & 5:34:31.9605,+22:00:52.148 \\
12228 & 2010-08-19 23:54:55 & 2.13 & 5:34:31.9700,+22:00:52.081 \\
13121 & 2010-10-06 04:32:35 & 1.59 & 5:34:31.9419,+22:00:52.046 \\
14686 & 2013-03-10 08:07:55 & 20.01 & 5:34:31.9227,+22:00:52.199 \\
16246 & 2013-10-22 12:26:45 & 20.01 & 5:34:31.9590,+22:00:52.764 \\
16247 & 2013-10-24 00:14:38 & 20.01 & 5:34:31.9737,+22:00:52.123 \\
    \hline
    \multicolumn{4}{l}{$a:$ \url{https://doi.org/10.25574/cdc.448}}
    \end{tabular}
\end{table}

The Crab nebula has been observed with the {\sl Chandra} X-ray Observatory ({\sl Chandra}) several times with both the ACIS and HRC instruments.  ACIS is a CCD imager with relatively large readout frame times even in continuous clocking mode ($2.8$~ms), and the HRC is a microchannel plate detector with timing resolution of $\sim$15.6~$\mu$s.  The HRC is therefore the appropriate detector with which to observe the Crab pulsar.  However, there is a timing error in the electronics that assigns the photon arrival time to the following event.  This is mitigated by running the detector in a mode where all events are telemetered down, and the timing is corrected during pipeline processing.  This mode can result in telemetry saturation due to the large number of events that can be recorded, so the Crab is often observed with the Low-Energy Transmission Grating (LETG) inserted as a neutral blocking filter, and with the lower level discriminator to detect events on the detector set higher, so only events with large signals are recorded.  We have analyzed a subset of the HRC-S/LETG observations (see Table~\ref{tab:chandraHRC}) here.  The HRC is not subject to pileup effects as in CCDs, but due to deadtime effects the number of counts detected becomes nonlinear at large count rates, such as at the peaks of the pulses.

All the data are processed using the CIAO analysis software \citep[v4.17][]{Fruscione2006} to apply the latest calibration, and the times are corrected for barycentric offsets using the CIAO tool {\tt axbary}, using the JPL-DE405 solar system ephemeris.  {\sl Chandra}'s spatial resolution is sufficient to isolate the pulsar as a point-like source within the diffuse nebula.  We have estimated the position of the pulsar by centroiding on this point-like source and collect source photons within a circle of radius $1.5$~arcsec.

\subsection{Einstein Probe/FXT}

EP/FXT timing mode observations of Crab are summarized in Table,\ref{tab:obslog}. The observational data were processed using the FXT Data Analysis Software (FXTDAS) version 1.20, alongside the FXT calibration database (CALDB) version 1.20. The data reduction pipeline, denoted as \textit{fxtchain}, was employed with precise source localization parameters: Right Ascension (R.A.) = 05h34m31.972s and Declination (Dec) = $22^{\circ}00^{'}52.07^{''}$. The standard Good Time Interval (GTI) selection criteria, ``\texttt{ELV>5\&\&COR>6\&\&SAA==0\&\&DYE\_ELV>30}'',  were applied. The events within the energy range of 0.5–10 keV were selected. The \textit{fxtbary} tool was utilized to convert the time of each event to barycentric corrected time using ephemeris ``de430\_plus\_MarsPC.bsp''.

\begin{table}[htb]
    \caption{Timing Mode Observation Log of Crab with EP/FXT}
    \centering
    \begin{tabular}{ccccc}
    \hline
    obsid          &   obs. date (UTC) & MJD   &  exposure \\
    \hline
    08500000012     &  2024-02-28       & 60368 & 2.6ks\\
    13600005102     &  2024-03-08       & 60377 & 8.3ks\\
    11900008151     &  2024-09-23       & 60576 & 5.9ks\\
    11900008152     &  2024-09-23       & 60576 & 5.9ks\\
    08500000313     &  2025-03-22       & 60756 & 2.0ks\\
    08500000314     &  2025-03-22       & 60756 & 2.9ks\\
    08500000324     &  2025-03-24       & 60758 & 3.0ks\\
    \hline
    \end{tabular}
    \label{tab:obslog}
\end{table}

\subsection{Fermi}
We queried the Fermi archive\footnote{\url{https://fermi.gsfc.nasa.gov/cgi-bin/ssc/LAT/LATDataQuery.cgi}} for all existing observations of the Crab pulsars up to November 5th, 2024 and the relevant spacecraft solutions. The data returned by this query were already usable for analysis without further cleaning, due to the strong signal from the Crab pulsar.
We then used the \texttt{gtbary} tool distributed with the \texttt{conda} package \texttt{fermitools}, to barycenter the data through the command
\texttt{gtbary file\_PH00.fits file\_SC00.fits file\_PH00\_bary.fits 83.633218 22.01446361111111}.

\subsection{Hitomi} 
We retrieved the final archival Crab data products for Hitomi from HEASARC.
For the Soft X-ray Spectrometer (SXS), we utilized the cleaned event FITS file to extract the high- and medium-resolution grades (Hp, Mp, and Ms; \citet{2018JATIS...4a1217I}) by selecting "ITYPE=0:2" within \texttt{xselect}.
The Hard X-ray Imager (HXI) data was obtained from the standard cleaned event FITS file, and extracted using a sky region around the Crab pulsar and extending up to a radius of 70 arcseconds from the image centroid.
Only one unit of the Soft Gamma-ray Detector (SGD) is available, because the observation taking place during the commissioning phase.
To increase the photon statistics, we selected photo-absorption event from the unscreened event FITS file, following the guidelines in Appendix 2 of \citet{2018PASJ...70...15H}.

Subsequently, barycentric correction was applied on the DE200 solar system ephemeris for the event FITS files using the command outlined in \citet{2018JATIS...4a1206T}:

\texttt{barycen orbext=ORBIT ra=83.633218 dec=+22.014464}.

\noindent
It should be noted that no energy selection was performed on the events; the approximate energy bands for the SXS, HXI, and SGD-1 photo-absorption events ranged from 2–10 keV, 2–80 keV, and 10–300 keV, respectively.

\subsection{HXMT}

The Crab observations from Insight-HXMT were processed using the official data pipeline \textit{hpipeline}, which performs filtering and calibration\footnote{Detailed documentation is available at \url{http://hxmten.ihep.ac.cn/}}. After these steps, photons within specific energy ranges were selected: HE: 25--250 keV, ME: 10--30 keV, and LE: 1--10 keV. The arrival times of the cleaned events were then corrected to the Solar System barycenter using the barycentric correction tool \textit{hxbary}, with the JPL ephemeris DE430 and source coordinates RA(J2000) = 05h34m31.972s, Dec(J2000) = +22$^\circ$00$^\prime$ 52.07\arcsec.

\subsection{IXPE}
We retrieved the IXPE data from the HEASARC archive, selecting all observations covering the Crab pulsar position. We did not run the mission pipeline, and trusted the cleaned event files from the official preprocessing done by the science operations center, as there should be no significant issue with the time stamps of the data. We barycentered the clean event data using the \texttt{barycorr} FTOOL, with the ICRS source position of (RA, Dec) = (83.63311445608998, 22.01448713834) and the DE440 solar system ephemeris.
To eliminate part of the contamination from the nebula, we applied a spatial filter to the event files, excluding events outside a 20$^{\prime\prime}$ circular region centered on the Crab pulsar. The astrometry showed some discrepancies in the order of 10-30 arcseconds between ObsIDs (but very consistent between different detector in the same observation), therefore we adjusted the source position manually.

\subsection{MAXI/GSC}

The MAXI/Gas Slit Camera (GSC) has been monitoring the Crab nebula continuously for over 16 years, since August 2009 \citep{moriiMAXIGSCMonitoring2011,sugizakiInOrbitPerformanceMAXI2011}. We retrieved the GSC data using the FTOOL \texttt{mxdownload\_wget} and generated daily cleaned event files with \texttt{mxproduct}, selecting events within a $1.6^\circ$ radius around the Crab\footnote{For the details of MAXI analysis, see \url{https://darts.isas.jaxa.jp/missions/maxi/analysis/}}. We then extracted events in the 2--20 keV energy range using \texttt{xselect}.
Barycentric correction was applied using the FTOOL barycen, based on the DE-200 solar system ephemeris and the source position of (RA, Dec) = (83.633083, 22.014500). To ensure sufficient statistics, we excluded days with fewer than 1000 counts. As a result, we obtained event files for 4089 days, up to July 10, 2025.

\subsection{NICER}
We followed the default procedure described in the NICER HEASARC page\footnote{\url{https://heasarc.gsfc.nasa.gov/docs/nicer/analysis_threads/nicerl2/}}.
We retrieved all NICER observations listed in the \texttt{nicermastr} table with nonzero exposure, and ran the \texttt{nicerl2} pipeline on each observation directory with default parameters.
Some observations had too few counts for a TOA and got discarded. We ended up with 89 usable observations between MJD 57970--60734 (Aug 2017--February 2025).
We ran \texttt{barycorr} on the cleaned event data, using the ICRS position from Simbad.
We repeated the procedure without running the nicerl2 pipeline and found consistent results, as the HEASARC-distributed clean event files were already processed with a standard pipeline by the instrument team and only contain little additional noise compared to a fully processed dataset where all parameters are correctly tweaked. Recent issues like the light leak\footnote{See \url{https://heasarc.gsfc.nasa.gov/docs/nicer/analysis_threads/light-leak-overview/}} require a reprocessing with HEASOFT$>$6.35.2, but the issue mostly affects spectral data.

For the end-to-end validation in Figure~\ref{fig:TOA_comparison}, we applied three different barycenter corrections to the HEASARC-distributed cleaned event files, using the following parameters:
\begin{itemize}
\item DE200 ephemeris, (RA, Dec) = (83.633083, 22.014500) (FK5)
\item DE430 ephemeris, (RA, Dec) = (83.63311445608998, 22.01448713834) (ICRS)
\item DE440 ephemeris, (RA, Dec) = (83.63311445608998, 22.01448713834) (ICRS)
\end{itemize}

The procedure was run on sciserver \citep{taghizadeh-poppSciServerSciencePlatform2020}

\subsection{NinjaSat}
NinjaSat is an astronomical X-ray CubeSat observatory. The satellite has a 6U form factor (34 $\times$ 24 $\times$ 11 cm$^3$) and a mass of 8~kg. It was launched into a Sun-synchronous orbit at an altitude of 530~km in November 2023 and began scientific observations on February 23, 2024. NinjaSat is equipped with two non-imaging gas X-ray detectors (Gas Multiplier Counters; GMCs). Each GMC is a detector filled with xenon-based gas, featuring an effective area of 16~cm$^2$ at 6~keV and sensitivity in the 2--50~keV energy range. X-rays from a target source are guided into the detectors through a collimator with a field-of-view (FoV) of 2.1 degrees (FWHM) placed in front of a gas cell. Detailed descriptions of the satellite and detectors are provided in \citet{tamagawaNinjaSatAstronomicalXray2025}. Each GMC assigns a time tag to individual X-ray photons with the aid of GPS signals. The time counter in the GMC has a resolution of 61~$\mu$s. Further details on the time tagging with GPS are found in \citet{OtaVerificationTimingMeasurement}.
To produce cleaned event data of the Crab Nebula observations, following the method described in \citet{takedaNinjaSatMonitoringType2025}, non-X-ray background signals and electrical noise were removed based on the characteristics of the analog waveforms. Good time intervals of the dataset were defined as the periods during which high voltage was applied to the GMC, the Crab Nebula was in the FoV, and GPS-PPS signals were received at 1~s intervals. Based on onboard estimates by the Attitude Determination and Control Subsystem, the spacecraft orbital position and velocity in the J2000 coordinate system were calculated and used for barycentric correction. We ran the “barycen” command with coordinates of R.A. = 83.633218 and Dec. = 22.014464 on the DE200 solar system ephemeris.

\subsection{NuSTAR}

We downloaded all public Crab data from the \nustar archive  from the High Energy Astrophysics Science Archive Research Center (HEASARC) with exposure longer than 100\,s.

The NuSTAR Data Analysis Guide\footnote{\url{https://heasarc.gsfc.nasa.gov/docs/nustar/analysis/nustar_swguide.pdf}} describes in detail the following steps, we direct the reader to the guide for details.
We ran \texttt{nupipeline} with default parameters.
We selected event lists in SCIENCE (01) and \texttt{SCIENCE\_SC} (06) modes.
Mode-01 data are those corresponding to all instruments and star trackers working in nominal conditions, while Mode-06 have only some of the camera head units (CHUs) working, leading to astrometry errors by 1--2$\amin$.
This mode makes the PSF slightly more difficult to model, and for precise spectral analysis data from these intervals should be used with care.
However, the arrival times of the photons are recorded with the same precision as Mode-01 data, and so they can be used for our work without issues.

For a sample of observations, we followed the general procedure used to process Mode-06 data (e.g. \citealt{waltonSoftStateCygnus2016}), splitting the intervals with different CHU combinations through the \texttt{nusplitsc} tool.
From both Mode-01 and Mode-06 data, we selected data from a 70\asec region around the centroid of the PSF (which, again, can move from the expected position in Mode-06 data). The selection was done through Python scripts that analyzed the images produced in each data mode interval looking for the maximum of the PSF.
Again, a more precise modeling of the PSF is not needed in this case, being interested only on the event arrival times and a rough energy measurement for each photon (and not a precise spectral modeling with PSF corrections), and being the Crab by far the most luminous source in the field of view.
The whole procedure was automated through the \texttt{heasarc\_retrieve\_pipeline} script v.0.2 \citep{bachettiHeasarcRetrieve}, installed on Sciserver \citep{taghizadeh-poppSciServerSciencePlatform2020}.
We later verified by eye that the selected regions were at least roughly consistent with the expected PSF positions.

For all observations, given that the Crab pulsar and nebula are not resolved in \nustar and they dominate the emission in the region, we used directly the cleaned event files distributed by HEASARC. In observations where both methods were used, we verified that results were consistent within error bars, with a slight increase in the scatter of the TOAs when using the full data, as expected due to the higher contribution from the nebula and the background.
This allowed for a quicker processing, with no manual intervention (that is often required when doing source selection due to astrometric uncertainty) and, importantly, without the long processing times typically associated with \texttt{nupipeline}.
Finally, we ran \texttt{barycorr} to refer the photon arrival times to the solar system barycenter.
We selected the ICRS reference frame, the DE430 JPL ephemeris, and the position of the Crab ra=83.63311445609 degrees, dec=22.01448713834 degrees.
For all observations, we used the latest clock correction file available, that provides an absolute time precision of $\sim60\,\mu$s \citep{bachettiTimingCalibrationNuSTAR2021}.

\subsection{Suzaku} 
We downloaded the final archival products of public Crab data for Suzaku HXD PIN and GSO from HEASARC.
Subsequently, we performed barycentric correction on the cleaned event FITS files using the \texttt{aebarycen} tool, as described in \citet{2008PASJ...60S..25T}:

\texttt{aebarycen ra=83.633218 dec=+22.014464}.

\noindent
Note that \texttt{aebarycen} is the prototype code of the \texttt{barycen} tool in the HEAsoft package, which computes the barycentric time on the DE200 solar system ephemeris.

It has been recognized that the ground station equipment responsible for time assignment encountered issues from 2012 to 2014, resulting in the degradation of Suzaku's absolute time accuracy by up to approximately 3 milliseconds. This observational data has not been omitted from the timing performance evaluation presented in Table \ref{tab:summary}.

\subsection{Swift/XRT}

The dataset used includes all Crab pulsar observations from 2005 to 2022, conducted with the XRT in WT mode. Only observations with valid radio ephemerides and where the pulsar is within 5 arcminutes of the on-axis direction were considered. Data were retrieved from the HEASARC public archive, then calibrated, filtered, and screened using the XRTDAS package, part of the HEASoft 6.32.1 software release, with default processing parameters.

For each observation, source events with grades 0–2 were extracted from a rectangular region (40 pixels wide) centered on the pixel with the highest count. This region encompasses approximately 94\% of the Point Spread Function (PSF) of the Crab pulsar, which is located near the center of the Nebula. The event arrival times were referenced to the Solar System Barycenter (SSB) using \texttt{barycorr} with the Crab pulsar's coordinates from the Jodrell Bank Monthly Ephemeris (RA = 05h34m31.972s, Dec = $22^{\circ}00^\prime 52^{\prime\prime}.07$; \citealt{lyneTwentyThreeYearsCrab1993}), the JPL-DE200 solar system ephemeris. The latter also accounts for time drift during the observation, using the mission clockfile (\texttt{swclockcor20041120v160.fits}, updated to May 6, 2024), and the spacecraft orbit as specified in the ObsID-specific attorb file.

\subsection{XMM-Newton}

We discuss here only the processing of EPIC pn data taken in timing mode (resolution=0.03ms with a live time of 99.5\%) and burst mode (resolution 7 microsecs with a live time of 3.0\%). In Timing mode, the Crab pulsar, gives a count rate at which pile-up is significant, which has the effect of slightly distorting the pulse profile. While Timing mode still provides very useful data on the timing accuracy of the pn instrument, the burst mode data timings are more precise.

A small number of the observations suffer from issues with unrecognized time-jumps and these are either excluded or the event files are modified to remove or adjust the post-jump event times. A few observations are also affected by telemetry or operational issues and are excluded. At the current time (May 2025), there are 107 usable observations (46 and 61 in Timing and Burst modes, respectively). 98 had sufficient signal-to-noise ratio for the analysis.

Elements of the analysis of the XMM-Newton pn timing data exploit the XMM Science Analysis System (SAS) - for the analysis reported here, SAS version 20 was used (cf the current version, SAS 22).

The analysis comprises the following main steps:

First, the pn event data from the Observation Data File (ODF) set for each observation are processed via the standard pn reduction chain, epproc. This is run as:

\texttt{epproc timing=YES burst=YES srcra=83.633216667 srcdec=22.014463889 withsrccoords=yes}

The cleaned event lists are then barycentrically corrected using the SAS task, barycen, and the FK5 (R.A., Dec.) 05:34:31.972 +22:00:52.070:

\texttt{barycen withtable=yes table='bary.ds:EVENTS' timecolumn='TIME' withsrccoordinates=yes \\ srcra='83.633216667' srcdec='22.014463889' processgtis=yes time=0 ephemeris=DE200)}

As can be seen, for routine monitoring of the pn timing accuracy, we adopt the DE200 solar system ephemeris for consistency with the Jodrell Bank radio data.

Timing mode data are measured from a rectangular region centred on 35,101 with width and height of 12 and 100 respectively. Burst mode data are taken from a rectangle centred at 35,71.5 with width and height, 20 and 70 respectively. For both Timing and Burst mode, the

Background is taken within a rectangle centred at 6,64 with width and height, 5 ands 63.5 respectively.

Subsets of events are subsequently extracted for the source and background regions from the observation cleaned event list.


\subsection{XRISM}

We utilized private observations of the Crab with XRISM Resolve obtained during the performance verification phase and the first cycle of the guest observation program in 2024.
The processing and calibration versions of the dataset were specified as TLM2FITS= '004\_002.15Oct2023\_Build7.011', PROCVER = '03.00.011.008', and CALDBVER= 'gen20240315\_xtd20240315\_rsl20240315', with ground-based calibration parameters incorporated for timing assignment.
We then reprocessed the unscreened event file with the pre-public timing-coefficient CALDB based on in-orbit calibration and employed standard screening criteria to get cleaned event file.
High- and medium-resolution grades (Hp, Mp, and Ms) were extracted by applying "ITYPE=0:2" in \texttt{xselect} to the cleaned event file.
A barycentric correction was then conducted on the DE200 solar system ephemeris using the command prescribed in \citet{2025JATIS_XRISM_Time1}:

\texttt{barycen orbext=ORBIT ra=83.633218 dec=+22.014464}.

\subsection{XTE}
We obtained a list of all RXTE/PCA observations pointed less than 0.1 degrees away from the Crab pulsar by querying the \texttt{xtemaster} table at HEASARC, using the \texttt{astroquery.heasarc} module. We selected only observations containing data in science event modes.
For each of these observations, we downloaded locally the event data (\texttt{fname}) and the relevant spacecraft file (\texttt{orbit\_file}), and barycentered the event data using the command

\texttt{barycorr fname outfile ra=83.63311446 dec=22.01448714 ephem=JPLEPH.DE440 refframe=ICRS \\orbitfiles=orbit\_file clobber=yes}.

To save space, we did not save the full information of the barycentered event files, but only the arrival times of the photons and the PHA column that could be used to extract energy-selected event lists.
We did not perform an energy selection for this work.

We processed in total 330 observations between MJD 50205--55927 (1996--2012). The procedure was executed on the Fornax infrastructure at NASA\footnote{\url{https://pcos.gsfc.nasa.gov/Fornax/}} \citep{jaffeFornaxInitiative2024}.

\section{Example of re-fitted timing model}

We report below the result of \texttt{model\_200.compare(model\_405)} in PINT, where \texttt{model\_200} is the original JBE model from the CGRO solution, closest to the start of the observation (MJD 55427.9964794566), and the \texttt{model\_405} is the result of the re-fitting of the model using the position and ephemeris used to barycenter the data, following the procedure in Section~\ref{sec:jpleph}.
The parameters marked with asterisks are those that were fitted, while the others were kept fixed to the values in the original JBE model. The apparent large differences in F0, F1, and F2 (compared to the quoted uncertainties) are due to the very low scatter of simulated data (see the model TRES) compared to the deviations produced by the different reference frames and ephemerides.
TZRMJD was fixed to a TOA close to PEPOCH.

\begingroup
    \fontsize{8pt}{8pt}\selectfont
    \begin{verbatim}
PARAMETER          Crab_55427.9964794566_DE200.par     crab_12228_none.par   Diff_Sigma1   Diff_Sigma2
--------------     -------------------------------   ---------------------   -----------   -----------
PSR                                     J0534+2200              J0534+2200
EPHEM                                        DE200                   de405
CLOCK                                      TT(TAI)                 TT(TAI)
UNITS                                          TDB                     TDB
START                                      55409.0   55408.000000064643586
FINISH                                     55440.0   55441.000000165256875
DILATEFREQ                                   False                   False
DMDATA                                       False                   False
NTOA                                             0                     501
TRES                                         40.38     0.09930508316423779                             !
POSEPOCH                                   40706.0                 40706.0
PX                                             0.0                     0.0
RAJ                                 5h34m31.97232s          5h34m31.99992s
DECJ                                 22d00m52.069s          22d00m52.0992s
PMRA                                           0.0                     0.0
PMDEC                                          0.0                     0.0
F0                              29.721114704239(0)     29.7211147070001(4)                     6878.81 ! *
PEPOCH                                     55424.0                 55424.0
F1                   -3.7117799999999996(0)×10E-18  -3.711789115(4)×10E-18                    -2055.94 ! *
F2                    2.4400000000000004(0)×10E-20      2.42957(18)×10E-20                      -57.71 ! *
PLANET_SHAPIRO                               False                   False
DM                                             0.0                     0.0
TZRMJD                             55424.000000163   55424.037999928536006                             !
TZRSITE                                          0                       0
TZRFRQ                                         inf                     inf
PHOFF                                       0.0(0)          0.0(2.0)×10E-7             nan          0.00 *
CHI2                                       Missing      494.06112706703493
CHI2R                                      Missing      0.9960909819899897
SEPARATION                         0.385001 arcsec
\end{verbatim}
\endgroup


\bibliography{iachec_crab}{}
\bibliographystyle{aasjournalv7}



\end{document}